\documentclass[11pt]{article}

\usepackage[left=1in,top=1in,right=1in,bottom=1in,head=.1in,nofoot]{geometry}

\usepackage{setspace,url,bm,amsmath} % For double-spacing, URL font, math symbols

\usepackage{titlesec} % Section header formatting
\titlelabel{\thetitle.\quad} % Section header formatting
\usepackage[margin=20pt]{subcaption}

\usepackage{amsmath,amsfonts,amsthm,amssymb}
\allowdisplaybreaks
\usepackage{graphicx}

\usepackage[colorlinks=true,linkcolor=black,citecolor=blue,urlcolor=blue]{hyperref}
\usepackage{threeparttable}
\usepackage{manyfoot}%
\usepackage{booktabs}%
\usepackage{algorithm}%
\usepackage{algorithmicx}%
\usepackage{algpseudocode}%
\usepackage{listings}%
\usepackage{dsfont}%
\usepackage{bm}
\usepackage{etoolbox}
\usepackage{natbib}
\usepackage{adjustbox}
\usepackage{threeparttable}
\usepackage{dcolumn}

\theoremstyle{definition}
\newtheorem{assumption}{Assumption}[section]
\newtheorem*{theorem*}{Theorem}
\newtheorem{theorem}{Theorem}[section]

\newtheorem{lemma}{Lemma}[section]
\newtheorem{remark}{Remark}[section]

\newtheorem*{corollary*}{Corollary}

\newtheorem{proposition}{Proposition}[section]

\DeclareFontFamily{U}{mathx}{\hyphenchar\font45}
\DeclareFontShape{U}{mathx}{m}{n}{
      <5> <6> <7> <8> <9> <10>
      <10.95> <12> <14.4> <17.28> <20.74> <24.88>
      mathx10
      }{}
\DeclareSymbolFont{mathx}{U}{mathx}{m}{n}
\DeclareFontSubstitution{U}{mathx}{m}{n}
\DeclareMathAccent{\widecheck}{0}{mathx}{"71}
\DeclareMathAccent{\wideparen}{0}{mathx}{"75}

\allowdisplaybreaks[4]

\def\T{\text{T}}
\def\Var{\text{Var}}
\def\Cov{\text{Cov}}

\def\v{{\varepsilon}}
\def\pr{P}

\title{\vspace{-1.5cm}\bf  On the achievability of efficiency bounds for covariate-adjusted response-adaptive randomization}
\date{}

\author{
\small
{
Jiahui Xin, \ \ Wei Ma\thanks{\small{Correspondence: \texttt{mawei@ruc.edu.cn}}}
}
\\
{\small Institute of Statistics and Big Data, Renmin University of China, Beijing, China}
}

\begin{document}
\doublespacing

\maketitle
\vspace{-1cm}
% , with the hope of improving treatment effect estimation efficiency

\begin{abstract}
In the context of precision medicine, covariate-adjusted response-adaptive randomization (CARA) has garnered much attention from both academia and industry due to its benefits in providing ethical and tailored treatment assignments based on patients’ profiles while still preserving favorable statistical properties. Recent years have seen substantial progress in understanding the inference for various adaptive experimental designs. In particular, research has focused on two important perspectives: how to obtain robust inference in the presence of model misspecification, and what the smallest variance, i.e., the efficiency bound, an estimator can achieve. Notably, \citet{armstrong2022asymptotic} derived the asymptotic efficiency bound for any randomization procedure that assigns treatments depending on covariates and accrued responses, thus including CARA, among others. However, to the best of our knowledge, no existing literature has addressed whether and how the asymptotic efficiency bound can be achieved under CARA. In this paper, by connecting two strands of literature on adaptive randomization, namely robust inference and efficiency bound, we provide a definitive answer to this question for an important practical scenario where only discrete covariates are observed and used to form stratification. Specifically, we consider a special type of CARA that separately implements doubly-adaptive biased coin design, a flexible and useful response-adaptive randomization procedure, within each stratum. For this kind of CARA, we prove that the stratified difference-in-means estimator achieves \citet{armstrong2022asymptotic}’s asymptotic efficiency bound, with possible ethical constraints on treatment assignments. Our work provides new insights and demonstrates the potential for more research regarding the design and analysis of CARA that maximizes efficiency while adhering to ethical considerations. Future studies could explore how to achieve the asymptotic efficiency bound for general CARA with continuous covariates, which remains an open question.

\vspace{12pt}
\end{abstract}

\section{Introduction}

Precision medicine can offer more individualized treatment plans based on patient characteristics. Traditional clinical trial designs often fail to meet the needs of precision medicine. Regulatory agencies such as the European Medicines Agency \citep{EMA2015} and the U.S. Food and Drug Administration \citep{FDA2023} have focused extensively on this issue and released guidelines on the incorporation of covariates. The key to designing new types of clinical trials in the context of precision medicine is how to simultaneously utilize both the patient's response to treatment and covariate information such as biomarkers.

The importance of incorporating covariates in clinical trials was recognized in early investigations such as deterministic
minimization proposed by \citet{taves1974minimization} and randomization-based minimization later introduced by \citet{Pocock1975}. \citet{efron1980randomizing} extended his earlier proposed biased coin design \citep{Efron1971} to account for covariates. For a comprehensive review, refer to \citet{rosenberger2008handling}. Nonetheless, they did not account for the responses of patients enrolled early in sequential trials. From the philosophy of assigning more patients to the treatment arm with better response, response-adaptive randomization (RAR) was developed \citep{Hu2006, Rosenberger2015}. Notably, Thompson sampling, proposed by \citet{thompson1933likelihood} from a Bayesian perspective, pioneered RAR and other online experimentation methods. RAR primarily includes urn models, such as the play-the-winner rule (deterministic version by \citet{zelen1969play}, later developed by \citet{Wei1978} into a randomized version), as well as methods based on sequential estimation \citep{melfi1998variablility, Hu2004}. For a recent review of the various types of RAR and their taxonomy, refer to \citet{robertson2023response}. Among these, the doubly-adaptive biased coin design (DBCD) \citep{eisele1994doubly, eisele1995doubly, Hu2004} has gained widespread application and attention in the past decades. DBCD can target any desired allocation proportions, including various allocation strategies such as the urn allocation \citep{rosenberger2004maximizing}, Neyman allocation \citep{neyman1934two}, and RSIHR allocation \citep{rosenberger2001}. Furthermore, \citet{hu2003optimality} demonstrated that DBCD achieves a smaller variance of the actual proportion and thus higher statistical power for a given allocation compared to other RAR, such as the sequential maximum likelihood
estimation (SMLE) procedure
and the
randomized play-the-winner (RPW) rule. Researchers have established a relatively complete theoretical framework for RAR under parametric model assumptions. Most recently, \citet{ye2024robustness} explored the theoretical properties of DBCD and the subsequent inference in the presence of model misspecification.

Covariate-adjusted response-adaptive randomization (CARA) \citep{Rosenberger2015, sverdlov2015modern} goes a step further by considering individual differences among patients, using information on the subjects' response variables and covariates to allocate them to better treatment groups with a higher probability. More specifically, the allocation depends on the allocation history, response history, covariate history, as well as the covariate of the incoming patient during the CARA process. Under parametric model assumptions, such as the logistic model in \citet{rosenberger2001covariate} and generalized linear models in \citet{zhang2007asymptotic}, there has been some theoretical progress on CARA. More recently, \citet{hu2015unified} proposed a unified framework for CARA, prioritizing both efficiency and ethical considerations. This unified family introduced new and desirable CARA designs, including examples of stratified adaptive designs. Furthermore, \citet{zhu2023covariate} employed semiparametric methods to sequentially estimate parameters and update allocation probabilities. The theoretical properties in \citet{hu2015unified} and \citet{zhu2023covariate} were derived under stratified randomization, which remains one of the most commonly used randomization methods in clinical trials \citep{Lin2015, Ciolino2019}.
This paper focuses on the scenario where only discrete covariates (or biomarkers) are observed. Specifically, we explore the CARA procedure that first stratifies subjects and then separately implements DBCD within each stratum. Implementing separate RAR designs within each stratum is conceptually straightforward \citep{Rosenberger2015}. This approach has been practically applied in clinical trials, such as the fluoxetine trial by \citet{tamura1994case} and the Bayesian adaptive biomarker-stratified phase II randomized design proposed by \citet{park2024bayesian}. Through stratification, we allow optimal allocations to vary across strata, aligning with the principles of precision medicine.

Literature has seen much attention in adaptive randomization, with several studies focusing on the enhancement of the efficiency of estimators and test statistics (e.g., \citealp{bugni2018inference, Bugni2019, ma2022regression, zhu2023covariate, tu2024unified, ye2024robustness}). In the identically and independently distributed (i.i.d.) case, \citet{Hahn1998} showed the semiparametric efficiency bound and Neyman allocation \citep{neyman1934two} further minimizes the variance. \citet{armstrong2022asymptotic} demonstrated that the optimized \citet{Hahn1998}'s bound cannot be improved in a much broader class of experimental designs, including various adaptive randomization procedures such as covariate-adaptive randomization (CAR), RAR, and CARA. Besides Neyman allocation based on the unconstrained optimization, \citet{armstrong2022asymptotic} also considered the optimal allocation based on the constrained optimization, which is the typical scenario considered under CARA. In CAR and “finely stratified” experiments, where units are matched within fixed-size blocks (e.g., pairs or quartets) and uniform randomization is carried out within each block, \citet{Rafi2023} and \citet{bai2023efficiency} derived the semiparametric efficiency bound and the corresponding efficient estimator, respectively. Nonetheless, both \citet{bai2023efficiency} and \citet{Rafi2023} assumed prespecified allocation probabilities in randomization, resulting in higher bounds than those proposed by \citet{armstrong2022asymptotic}; however, the requirement of prespecified allocation probabilities is not necessary under CARA.

Therefore, ambiguity still persists regarding whether and how the asymptotic efficiency bounds in \citet{armstrong2022asymptotic} can be achieved for CARA. By connecting two strands of literature---robust inference and efficiency bounds---we provide definitive answers for an important practical scenario when only discrete covariates are observed and used for stratification. Specifically, for a special type of CARA that separately implements DBCD within each stratum, the stratified difference-in-means estimator achieves \citet{armstrong2022asymptotic}'s bound with constrained optimization. In other words, the stratified difference-in-means estimator under CARA attains the minimal asymptotic variance in the class of randomization considered in \citet{armstrong2022asymptotic}, including all adaptive randomization with discrete covariates. To establish the results, we utilize techniques for both design and inference from recent literature such as \citet{Bugni2019}, \citet{Bai2022}, and \citet{ye2024robustness}. Our results highlight the potential for more research regarding CARA that maximizes efficiency while adhering to ethical considerations, although it remains unresolved how to achieve the asymptotic efficiency bound for general CARA with continuous covariates.

The organization of this paper is as follows. In Section 2, we introduce our framework, including the setup and optimal allocation based on optimization. Section 3 reviews the key results from \citet{armstrong2022asymptotic} and presents the asymptotic efficiency bound for CARA. In Section 4, we prove that the stratified difference-in-means estimator is the asymptotically efficient estimator for an important class of CARA. Section 5 provides the numerical results, and conclusions are drawn in Section 6. Appendix contains the proof of main results, the auxiliary results and the additional simulations. The R and C++ code used to generate the presented results, is available on GitHub at \href{https://github.com/jiahui-xin/cara-bound}{https://github.com/jiahui-xin/cara-bound}.

\section{Framework}
\subsection{Setup}
We consider a setting in which baseline covariates $X_i$ are discrete and potential outcomes $\{Y_i(w)\}_{w\in\mathcal W}$ are associated with subject $i$, where $\mathcal W=\{0,1\}$ is the set of two possible treatment assignments. We assume that $X_i\in\mathcal X,\{Y_i(w)\}_{w\in\mathcal W}\in\mathbb R$ are i.i.d. from some population, where $\mathcal X$ is a discrete set.
In the randomization procedure, the researcher chooses a treatment assignment $W_i$ for each subject $i$, and observes $X_i$ and $Y_i=Y_i(W_i)$ for each subject $i$. The series of assignments is denoted by $W^{(n)}=(W_{1},\dots,W_{n})$. When $W_i$ is i.i.d., then it leads to the standard (i.e., non-adaptive) case. Within a CARA framework, the treatment assignment for the $i$-th subject $W_i$ depends on the initial $i$ baseline covariates $X^{(i)}=(X_1,\dots,X_{i})$ which includes the covariate of the incoming patient $X_i$, as well as the response history $Y^{(i-1)}=(Y_{1},\dots,Y_{i-1})$ and assignment history $W^{(i-1)}=(W_{1},\dots,W_{i-1})$. If no additional baseline covariates are observed, we let $X=0$. Denote $\mu(x,w)=E\{Y_i(w)|X_i=x\},\sigma^2(x,w)=\Var\{Y_i(w)|X_i=x\}$ and $\mu(w)=E\{Y_i(w)\},\sigma^2(w)=\Var\{Y_i(w)\}$. Denote $p(x)=\pr(X_i=x)$,  $n(x)=\sum_{i=1}^n I\{X_i=x\}$ and $n(x,w)=\sum_{i=1}^n I\{X_i=x,W_i=w\}$, where $I\{\cdot\}$ is the indicator function. 

Based on the observed data, the researcher forms an estimator for the average treatment effect (ATE), denoted by $\tau=\mu(1)-\mu(0)$. Substantial literature has been dedicated to seeking more efficient estimation methods, and this paper specifically focuses on semiparametric efficiency in the sense of \citet{van1998asymptotic}. %Semiparametric efficiency bounds assert that no uniform efficiency improvement is possible over a class of distributions rich enough to include a particular one-dimensional submodel, known as the ``least favorable submodel."
To be more specific, the efficiency bound is a lower bound on the asymptotic variance of any well-behaved, or “regular,” estimators of ATE \citep{van1998asymptotic}. If the asymptotic variance of an estimator matches this bound, then no uniform efficiency improvement is achievable. In the class broader than CARA, \citet{armstrong2022asymptotic} examined the same efficiency bound, but optimized it as in the i.i.d. case \citep{Hahn1998}.
\begin{remark}
    The response-adaptive randomization literature such as \citet{hu2003optimality} and \citet{hu2006asymptotically} used  ``efficiency" to indicate maximizing the power of the Wald's test under a specified parametric model, especially, with binary response and no observed covariate. To clarify the term ``efficiency" used in this paper, our objective is to minimize the asymptotic variance of ATE estimator under semiparametric model.
\end{remark}
\subsection{Allocation based on optimization}

%Denote the (conditional) propensity $e(x)=P(W_i=1|X_i=x)$ and the marginal propensity $e=P(W_i=1)$. 
In the i.i.d. case, \citet{Hahn1998} gave the semiparametric efficiency bound for a given propensity $e(x)=\pr(W_i=1|X_i=x)$,
\begin{align}
\label{bound_no_cnstrt}
    v_{e(\cdot)}=\Var\{\mu(X_i,1)-\mu(X_i,0)\}+E\left\{\frac{\sigma^2(X_i,0)}{1-e(X_i)}\right\}+E\left\{\frac{\sigma^2(X_i,1)}{e(X_i)}\right\}.
\end{align}
To further minimize the bound in (\ref{bound_no_cnstrt}), we can choose $e^*(\cdot)$ as Neyman allocation \citep{neyman1934two} such that
\begin{align}
\label{neyman_alloc}
\frac{\sigma^2(X_i,0)}{\{1-e^*(X_i)\}^2}=\frac{\sigma^2(X_i,1)}{e^*(X_i)^2},
\end{align}
which is obtained by solving the unconstrained optimization problem
\[\min_{e(\cdot)} v_{e(\cdot)}.\] If there are no observed $X$, we denote the bound as $V_e$.

In clinical trials, constraints may arise due to ethical concerns or budget restrictions (for more discussions, see \citet{Hu2006}). Consider a constrained optimization problem for the propensity $\pi(x)$.
{\small \begin{eqnarray}
\label{bound_cnstrt}
\nonumber\mathrm{min}_{\pi(\cdot)}&\ v_{\pi(\cdot)}= \Var\{\mu(X_i,1)-\mu(X_i,0)\}+E\left\{\frac{\sigma^2(X_i,0)}{1-\pi(X_i)}\right\}+E\left\{\frac{\sigma^2(X_i,1)}{\pi(X_i)}\right\},\\
\text{subject to}& E_{\theta^*}\left[r(X_i,1)\pi(X_i) + r(X_i,0)\{1-\pi(X_i)\}\right]\leq c,
\end{eqnarray}
}
where $c$ can be a vector and $r(\cdot)$ is a vector-valued function with the same dimension. Let $\pi^*(X_i)$ be the solution of optimization (\ref{bound_cnstrt}). If there are no observed $X$, we denote the bound as $v_\pi$.
\begin{remark}(Optimal allocation without constraints)
By setting the constraint border ${c}$ to $\infty$ in (\ref{bound_cnstrt}), the solution of the optimization is Neyman allocation. 
\end{remark}
\begin{remark}(Optimal allocation with constraints)
    Target allocation is a pivotal component of DBCD. Conventionally, various optimal allocation rules have been derived from optimization under certain model or distributional assumptions \citep{Hu2006}. In this paper, we directly solve the optimization problem (\ref{bound_cnstrt}) without making these assumptions.
    Note that if we define \begin{equation*}
        r(X_i,w)=\left(\frac{\mu(x,w)I\{X_i=x\}}{\pr(X_i=x)}\right)_{x\in\mathcal X}\in\mathbb R^{|\mathcal X|}
    \end{equation*}
    and let ${c}\in\mathbb R^{|\mathcal{X}|}$
    in the constrained optimization \ref{bound_cnstrt}, then the solution is 
\begin{equation}
\rho_x=\rho(\theta_x)
=\left\{\begin{array}{l}
\frac{c_x-\mu(x,0)}{\mu(x,1)-\mu(x,0)}, \quad s_x=1, \\
\frac{\sigma(x,1)}{\sigma(x,1)+\sigma(x,0)}, \quad \text { otherwise. }
\end{array}\right.
\end{equation}
where ${\theta}_x=\left(\mu(x,1), \sigma^2(x,1), \mu(x,0), \sigma^2(x,0)\right)$. Let $e^{\star}(x)=\sigma(x,1) / \{\sigma(x,1)+\sigma(x,0)\}$, then $s_x$ is set as
$$
s_x= \begin{cases}1, & e^*(x)\mu(x,1)+\{1-e^*(x)\}\mu(x,0)>c_x, \\ 0, & \text { otherwise. }\end{cases}
$$
By modifying the constraint border $c_x$, various allocations $\rho_x$ can be obtained. Specifically, if $c_x$ is relatively large such that $s_x=0$, $\rho_x$ will be Neyman allocation. If   $c_x$ is set as 
\begin{align*}
\mu(x,0)+\frac{\{\mu(x,1)-\mu(x,0)\}\sigma(x,1) \sqrt{\mu(x,0)}}{\sigma(x,1) \sqrt{\mu(x,0)}+\sigma(x,0) \sqrt{\mu(x,1)}}
\end{align*}
and at the same time
$
e^*(x)\mu(x,1)+\{1-e^*(x)\}\mu(x,0)>c_x,
$
then we have \[\rho_x=\frac{\sigma(x,1) \sqrt{\mu(x,0)}}{\sigma(x,1) \sqrt{\mu(x,0)}+\sigma(x,0) \sqrt{\mu(x,1)}},\]which is RSIHR allocation \citep[p. 13]{rosenberger2001,Hu2006}.
\end{remark}

\section{The asymptotic efficiency bound for CARA}
%We consider a finite-dimensional parametric model indexed by $\theta$ and  are interested in efficiency bounds at a particular $\theta^*$. We use least favorable submodels to derive semiparametric efficiency bounds in infinite-dimensional settings. For clarity, we subscript expectations $E_\theta$ and probability statements $P_\theta$ by $\theta$ to indicate that data is drawn from this model.

In this section, we review some key results from \citet{armstrong2022asymptotic} and provide the efficiency bounds. 

For a regular parametric submodel indexed by $\theta$, denote $f_X(x;\theta)$ and $f_{Y(w)|X}(y|x;\theta)$ as the probability density function of $X_i$ and $Y_i(w)|X_i$, respectively.  {Within a CARA framework, $W_i=w_i(X^{(i)},Y^{(i-1)},U)$ is a measurable function of $(X^{(i)},Y^{(i-1)},U)$ and $U$ is an exogenous random variable independent of the sample, which allows for randomization.} Denote $p_U$ as the density of the exogenous randomness $U$ which does not depend on $\theta$. Then, the likelihood of $U,X_1,\dots,X_n,Y_{1},\dots,Y_{n}$ can be written as
\begin{align}
\label{likelihood}
    p_U(u)\prod_{i=1}^n\left\{f_X(x_i;\theta)\prod_{w\in\mathcal W}f_{Y(w)|X}(y_i|x_i;\theta)^{I\{W_i=w\}}\right\},
  \end{align}
where $W_i=W_i(x_1,\dots,x_i,y_1,\dots,y_{i-1},u)$.

{%Because the treatment rule is determined once the exogenous randomness $U$ is given, we can take the observed data to be $X_1,\dots,X_n,Y_1,\dots,Y_n$ and $U$, so that the likelihood is given by (\ref{likelihood}).

}
Under the assumptions of quadratic mean differentiability and pathwise differentiability of the ATE parameter within the submodel, as defined by \citet{van1998asymptotic}, \citet{armstrong2022asymptotic} demonstrated the same least favorable submodel as in the i.i.d. case studied by \citet{Hahn1998}, but with optimal allocations. Building on this result, he further established the efficiency bound.

\subsection{Optimal allocation without constraints (Neyman allocation)}

The following proposition establishes the efficiency bound in the absence of constraints, which requires the use of the optimal (Neyman) allocation.
\begin{proposition}\citep{armstrong2022asymptotic}
\label{prop1}
    Assume some regularity conditions on the submodel. Let $\widehat{ATE}_n=\widehat{ATE}_n(X^{(n)},Y^{(n)},W^{(n)})$ by any sequence of regular estimator under any sequence of treatment rule $W_i(X^{(n)},Y^{(i-1)},U)$. For any loss function $L$ that is subconvex, we have
    \begin{align*}
        \sup_A\liminf_{n\to\infty}\sup_{h\in A}E_{\theta^*+\frac{h}{\sqrt{n}} }L(\sqrt n \{\widehat{ATE}_n-ATE(\theta^*+\frac{h}{\sqrt{n}})\})
        \geq E_{T\sim \mathcal N(0,v_{e^*(\cdot)})}L(T),
    \end{align*}
    where the first supremum is over all finite sets in $\mathbb R$ and $e^*(\cdot)$ is given by Neyman allocation (\ref{bound_no_cnstrt}). 
\end{proposition}
\begin{proof}
    The result is from Corollary 4.1 in \citet{armstrong2022asymptotic}. 
\end{proof}
%By choosing one specific submodel, Proposition 1 establishes a lower bound of efficiency bound.
The results essentially mean that $v_{e^*()}$ is the lower bound on the asymptotic variance of all the regular estimators. The proposition holds for any sequence of treatment rules $W_i(X^{(n)},Y^{(i-1)},U)$ which includes, besides RAR, CARA, also CAR and matched-pair experiments \citep{Bai2022,bai2023efficiency}.

For the bounds with constraints, recall the constrained optimization problem (\ref{bound_cnstrt}). Consider any treatment rule $W_i(X^{(n)},Y^{(i-1)},U)$ that satisfies the constraint in (\ref{bound_cnstrt}) on average, in the sense that
\begin{align}
\label{bound_ave}
    \frac{1}{n}\sum_{i=1}^n\sum_{w\in\mathcal W}r(X_i,W_i)\leq c+o_{P_{\theta^*}}(1).
\end{align}

\subsection{Optimal allocation with constraints}
The proposition below presents the efficiency bound with constraints, where the optimal allocation can be derived by solving the corresponding optimization problem.
\begin{proposition} \citep{armstrong2022asymptotic}
\label{prop2}
    Assume some regularity conditions on the submodel. Let $\widehat{ATE}_n=\widehat{ATE}_n(X^{(n)},Y^{(n)},W^{(n)})$ by any sequence of regular estimator under any sequence of treatment rule $W_i(X^{(n)},Y^{(i-1)},U)$ satisfying (\ref{bound_ave}). For any loss function $L$ that is subconvex, we have
    \begin{align*}
    \sup_A\liminf_{n\to\infty}\sup_{h\in A}E_{\theta^*+\frac{h}{\sqrt{n}}}L(\sqrt n \{\widehat{ATE}_n-ATE(\theta^*+\frac{h}{\sqrt{n}})\})
        \geq E_{T\sim \mathcal N(0,v_{\pi^*(\cdot)})}L(T),
    \end{align*}
    where the first supremum is over all finite sets in $\mathbb R$ and $\pi^*(\cdot)$ is the solution of optimization problem (\ref{bound_cnstrt}).
\end{proposition}
{
\begin{proof}
    Here, we provide a sketch of the proof. For formal details, see Section \ref{secA1} in the Appendix.

   First we define $\pi(x,w)=\pr\{W_i=w|X_i=x\}$ for simplicity of deduction. Notice that optimization problem (\ref{bound_cnstrt}) is equivalent with
    {\small \begin{eqnarray*}
\nonumber\mathrm{min}_{\pi(\cdot)}&\ v_{\pi(\cdot)}= \Var\{\mu(X_i,1)-\mu(X_i,0)\}+E\left\{\frac{\sigma^2(X_i,0)}{\pi(X_i,0)}\right\}+E\left\{\frac{\sigma^2(X_i,1)}{\pi(X_i,1)}\right\},\\
\text{subject to}&\ \nonumber E_{\theta^*}\left[r(X_i,1)\pi(X_i,1) + r(X_i,0)\pi(X_i,0)\right]\leq c,\\
\text{and}&\ \pi(x,0)+\pi(x,1)=1 \text{ for all x},
\end{eqnarray*}
}
which is restricted and simplified version of (12) in \citet{armstrong2022asymptotic}. Our considered (\ref{bound_cnstrt}) restricts the case to the binary treatment with allocation probabilities adding up to one, as in \citet{hahn2011adaptive}.

Based on the above optimization problem, we let $\lambda(x)$ and $\eta$ denote the Lagrange multipliers for the constraint. Noting only the second term of $v_{\pi(\cdot)}$ depends on $\pi(\cdot)$, the Lagrangian can be written as
\begin{align*}
\mathcal L=&E\Big(\frac{\sigma^2(X_i,0)}{\pi(X_i,0)}+\frac{\sigma^2(X_i,1)}{\pi(X_i,1)}+\eta^\T\left[r(X_i,1)\pi(X_i,1) + r(X_i,0)\pi(X_i,0)\right]+\lambda(X_i)\{\pi(X_i,0)+\pi(X_i,1)-1\}\Big).
\end{align*}

Write down the first order condition of this optimization problem and plug it into the representation of $v_{\pi^*(\cdot)}$. Follow the proof details of Theorem 5.1 and Corollary 5.1 in \citet{armstrong2022asymptotic}, we can deduce the remaining part.
\end{proof}

\begin{remark}
    Proposition \ref{prop2} simplifies the case from \citet{armstrong2022asymptotic} because optimization problem (\ref{bound_cnstrt}) only considers an inequality constraint related to the outcome. % because optimization problem (\ref{bound_cnstrt}) considers only the outcome constraint without addressing the allocation probability constraint. 
    In particular,  \citet{armstrong2022asymptotic} also allows for allocation probabilities with a summand less than one.
\end{remark}

%By choosing one specific submodel, Proposition 2 establishes a lower bound of efficiency bound.
Proposition \ref{prop2} holds for any treatment rules $W_i(X^{(n)},Y^{(i-1)},U)$ satisfying (\ref{bound_ave}) which includes, besides RAR and CARA, among others. The proposition shows that $v_{\pi^*()}$ is the lower bound on the asymptotic variance of all the regular estimators. If Neyman allocation $e^*(\cdot)$ satisfies the constraints in (\ref{bound_cnstrt}), then we have $\pi^*(\cdot)=e^*(\cdot)$ and $v_{\pi^*(\cdot)}=v_{e^*(\cdot)}$. Otherwise, the optimal solution is on the border of constraints, and consequently, $\pi^*(\cdot)\ne e^*(\cdot)$ and $v_{\pi^*(\cdot)}>v_{e^*(\cdot)}$.

\section{The asymptotically efficient estimation under CARA}
In this section, we first introduce the theoretical properties of CARA based on optimization in Section 4.1. Secondly, we show in Section 4.2 that the stratified difference-in-means estimator under CARA achieves the efficiency bound given in Section 3.

\subsection{CARA based on optimization}
Within the stratum $X=x$ that contains $n(x)$ subjects, a DBCD procedure is implemented. For $w = 0, 1$, we define $n(x,w)$ as the number of subjects in treatment $w$ within stratum $X=x$, with $n(x,0) + n(x,1) = n(x)$.

In the implementation of DBCD, a design model needs to be specified. This model  assumes that $Y_i(w)|X_i=x$ follows a distribution represented by $h_{{\theta}_{w,x}}$, though it is not necessarily the true model. The distribution form of $h_{{\theta}_{w,x}}$ is known, where ${\theta}_{w,x} \in \Theta \subset \mathbb{R}^{d_w}$ represents the unknown parameters. The number of unknown parameters for $h_{{\theta}_{w,x}}$ is denoted as $d_{w,x}$. We assume that $\Theta$ is locally compact with a countable base, and $d_{1,x}=d_{0,x}=d$.

The DBCD procedure sequentially estimates the parameters by maximizing the likelihood function determined by $h_{{\theta}_{w,x}}$ and updates the allocation probability for the next patient. Based on the first $j$ subjects within the stratum $X=x$, $\widehat{{\theta}}_{j(w),x}$ represents the estimates of ${\theta}_{w,x}$ for treatment $w\in\{0,1\}$.

One objective of the allocation scheme using the design model is to ensure that $n(x,1) / n(x)$ tends to $\rho_x$, where $\rho_x=\rho({\theta}_x)=\rho\left({\theta}_{0,x}, {\theta}_{1,x}\right)$ denotes the optimal allocation within stratum $X=x$, derived from some optimization problem (\ref{bound_cnstrt}). Moreover, $g(x, y)$ is the allocation function from $[0,1] \times[0,1]$ to $[0,1]$. Throughout this paper, we consider the allocation function to be in the form proposed by \citet{Hu2004}:
\begin{equation}
\label{g_func}
g(x, y)=\left\{\begin{array}{l}
1, \quad  x=0, \\
0, \quad x=1, \\
\frac{y(y / x)^\gamma}{y(y / x)^\gamma+(1-y)\{(1-y) /(1-x)\}^\gamma}, \quad \text { otherwise, }
\end{array}\right.
\end{equation}
where $\gamma \geq 0$. Next, we elaborate on the procedure of CARA, which implements the DBCD within each stratum.%, recognizing that the design model $h_{{\theta}_{w,x}}$ may be misspecified.
\newline

(1) Start each stratum with a permuted block randomization of $2 n_0$ subjects, ensuring that $n_0 (\geq 2)$ subjects are allocated to each treatment group. 

(2) Within the stratum $X=x$, for the $(j+1)$th stage where $j \geq 2 n_0$, assign the $(j+1)$th subject to treatment 1 with probability $g\left(n(x,1) / j, \widehat{\rho}_{j,x}\right)$. Here $\widehat{\rho}_{j,x}=\rho\left(\widehat{{\theta}}_{j(1),x}, \widehat{{\theta}}_{j(0),x}\right)$.
\newline

\begin{remark}
    In step (1) of the DBCD procedure, we implement a permuted block randomization for each group within each stratum to establish a start-up design. Alternatively, one may choose another restricted randomization method or complete randomization with appropriate burn-in sample sizes, as suggested by \citet{haines2015start}.
\end{remark}

Below, we list the standard regularity conditions outlined in \citet{ye2024robustness} to ensure the robustness of stratified DBCD against design model misspecification.

\begin{assumption}
The response sequence $\left\{Y_i(w)\right\}_{i=1}^n$ is i.i.d. random variables for $w\in\{0,1\}$. Moreover, for some $\epsilon>0, E\left|Y_i(w)\right|^{2+\epsilon}<\infty$.
\end{assumption}
The second condition pertains to the target allocation proportion function $\rho({z})$.

\begin{assumption}For the proportion function ${z}=\left({z}_0, {z}_1\right)=\left(z_{01}, \ldots, z_{0 d}, z_{11}, \ldots, z_{1 d}\right) \rightarrow$ $\rho({z}): \mathbb{R}^{2 d} \rightarrow[0,1]$, there exists some $\delta>0$ such that
\begin{align*}
    \rho({z})=&\rho({\theta})+\sum_{k=0,1} \sum_{m=1}^d\left(z_{k m}-\theta_{k m}\right) \frac{\partial \rho}{\partial z_{k m}}\Big|_{{\theta}}+o\left(\|{z}-{\theta}\|^{1+\delta}\right), \text{\  as ${z} \rightarrow {\theta} $.}
\end{align*}
\end{assumption}
Further required assumptions are specified for the estimators $\widehat{{\theta}}_{n(x)(w),x}, w\in\{0,1\}, n \geq 2 n_0$.

\begin{assumption}For $w=0,1$ and $n(x) \geq 2 n_0$, $\widehat{{\theta}}_{n(x)(w),x}$ is the solution to the estimating equations given by
$$
\sum_{X_i=x,W_i=w} {\psi}_{{\theta}_w}\left(Y_i(W_i)\right)=0
$$
subject to the following conditions. Here, the expectation is taken over the conditional distribution of $Y_i(w)$ given $X_i=x$:

(5.1) For each ${\theta}_w \in \Theta, {\psi}_{{\theta}_w}(y)$ is some measurable function with values in $\mathbb{R}^d$, and ${\psi}_{{\theta}_w}(y)$ is separable: there is a probability null set $\mathcal{N}$ and a countable set $\Theta^{\prime} \subset \Theta$ such that for every open set $U \subset \Theta$ and every closed interval $A$, the sets $\left\{y: {\psi}_{{\theta}_w}(y) \in A, \forall {\theta}_w \in U\right\}$ and $\left\{y: {\psi}_{{\theta}_w}(y) \in A, \forall {\theta}_w \in U \cap \Theta^{\prime}\right\}$ differ by a subset of $\mathcal{N}$.

(5.2) $E\left\{\sup _{{\theta}_w^{\prime} \in U}\left|{\psi}_{{\theta}_w^{\prime}}(y)-{\psi}_{{\theta}_w}(y)\right|\right\} \rightarrow 0$ as the neighborhood $U$ of ${\theta}_w^{\prime}$ shrinks to $\left\{{\theta}_w\right\}$, where $|\cdot|$ is taken to be the sup-norm: $\left|{\theta}_w\right|=\max \left(\left|\theta_{w, 1}\right|, \ldots,\left|\theta_{w, d}\right|\right)$.

(5.3) The expected value $\Psi_w\left({\theta}_w\right)=E {\psi}_{{\theta}_w}(y)$ exists for all ${\theta}_w \in \Theta$, and has a unique zero at ${\theta}_w={\theta}_w^0$.

(5.4) There exists a continuous function which is bounded away from zero, $b_w\left({\theta}_w\right) \geq b_0>$ 0 , such that (i) $\sup _{{\theta}_w}\left\{\left|\psi_{{\theta}_w}(y)\right| / b_w\left({\theta}_w\right)\right\}$ is integrable, (ii) $\liminf _{{\theta}_w \rightarrow \infty}\left\{\left|\Psi_w\left({\theta}_w\right)\right| / b_w\left({\theta}_w\right)\right\} \geq 1$, (iii) $E \limsup _{{\theta}_w \rightarrow \infty}\left\{\left|{\psi}_{{\theta}_w}(y)-\Psi_w\left({\theta}_w\right)\right| / b_w\left({\theta}_w\right)\right\}<1$.
\end{assumption}

\begin{assumption}Define $$u\left(y, {\theta}_w, m\right)=\sup _{\left|\iota-{\theta}_w\right| \leq m}\left|{\psi}_\iota(y)-{\psi}_{{\theta}_w}(y)\right|.$$ For $k=0,1$, the following conditions are made, with the expectation taken over the conditional distribution of $Y_i(w)$ given $X_i=x$:

(6.1) For each ${\theta}_w \in \Theta,\left|{\psi}_{{\theta}_e}(y)\right|^2$ is integrable.

(6.2) $\Psi_w$ has a non-singular derivative $\Lambda_w$ at ${\theta}_w^0$.

(6.3) There exist positive numbers $a, b, c, m, \alpha, \beta$, and $m_0$ such that $\alpha \geq \beta>2$, and (i) $\left|\Psi_w\left({\theta}_w\right)\right| \geq a\left|{\theta}_w-{\theta}_w^0\right|$ for $\left|{\theta}_w-{\theta}_w^0\right| \leq m_0$, (ii) $E u\left(x, {\theta}_w, m\right) \leq b m$ for $\left|{\theta}_w-{\theta}_w^0\right|+m \leq m_0$, (iii) $E u^\alpha\left(x, {\theta}_w, m\right) \leq c m^\beta$ for $\left|{\theta}_w-{\theta}_w^0\right|+m \leq m_0$.

(6.4) $\partial \Psi_w\left({\theta}_w\right) / \partial {\theta}_w$ is Lipschitz in a neighborhood of ${\theta}_w^0$.
\end{assumption}

Many common estimators in clinical trials, such as moment estimators and maximum likelihood estimators, can be framed as solutions to the estimating equation described in Assumption 5, utilizing an appropriate estimating function ${\psi}_{{\theta}_{{w}}}$. With Assumptions 3-6, one can establish the consistency and thelaw of the iterated logarithm of $M$-estimators based on $\left\{Y_i(W_i)\right\}_{X_i=x,W_i=w}$. Furthermore, the law of large numbers for allocation proportions can also be established.

\setcounter{theorem}{0}
\begin{theorem}
\label{thm1}
    If Assumptions 1--4 are satisfied, then for $x\in\mathcal X$,
    \[\hat{{\theta}}_x\to{\theta}_x,\ \hat\rho_x\to\rho_x, \ n(x,1)/n(x)\to\rho_x \text{ a.s.}\]
    
\end{theorem}
\begin{proof}
Theorem 4 in \citet{ye2024robustness} considered the statistical properties of DBCD within a single stratum: conditional on $n(x)$, $
        \hat{{\theta}}_x\to{\theta}_x,\ \hat\rho_x\to\rho_x, \ n(x,1)/n(x)\to\rho_x \text{ a.s.}$
    From the superpopulation assumption, $n(x)\sim Multinomial (n,p(x))$ for $x\in\mathcal X$, and hence $
        n(x)/n\to p(x)\text{ a.s.}$ Since
    $n(x)$ uniformly goes to infinity a.s., we can obtain the result.
\end{proof}
\begin{remark}
    Theorem \ref{thm1} establishes the almost sure convergence of the parameter estimator, estimated proportion function, and actual proportion within each stratum. Based on this, we can deduce the overall convergence of each to its respective expected value, with the expectation taken over the covariate $X$. For example, ${\sum_{x\in\mathcal X}n(x,1)}/{n}\to E(\rho_X)$ a.s.
    %\[\hat{{\theta}}\to E({\theta}_X),\ \hat\rho\to E (\rho_X), \ \frac{\sum_{x\in\mathcal X}n(x,1)}{n}\to E(\rho_X) \text{ a.s.}\]
\end{remark}
\subsection{The asymptotic variance of stratified difference-in-means}
In this section, we study the inference under CARA with constraints. We aim to estimate the average treatment effect $\tau=E\left\{Y_i(1)-Y_i(0)\right\}$ based on the observed data $\left\{\left(Y_i, X_i, W_i\right)\right\}_{i=1}^n$.
Our focus is on the stratified difference-in-means, a widely used method for estimating treatment effects in randomized clinical trials, particularly under CAR \citep{Bugni2019,ma2022regression}.
The stratified difference-in-means estimator is defined as
\begin{equation*}
    \hat\tau=\sum_{x\in\mathcal X}\frac{n(x)}{n}\{\hat\mu(x,1)-\hat\mu(x,0)\},
\end{equation*}
where $\hat\mu(x,w)=\sum_{X_i=x,W_i=w}Y_i/n(x,w)$ is the average of responses in the treatment $w$ and stratum $x$. Next, we demonstrate the consistency and asymptotic normality of the stratified difference-in-means estimator $\hat\tau$ under CARA with constraints. The results show that the stratified difference-in-means estimator achieves the lower bound $v_{\pi^*(\cdot)}$ in Proposition \ref{prop2}.

\begin{theorem}
\label{thm2}
    If Assumptions 1--4 are satisfied, then 
    we have $\hat\tau \rightarrow \tau$ a.s., and $\sqrt{n}(\hat\tau-\tau) \xrightarrow{d} \mathcal N\left(0, \sigma_{\hat\tau}^2\right)$, where
\begin{align*}
    \sigma_{\hat\tau}^2=\sum_{x\in\mathcal X}p(x)\Big\{\frac{1}{\rho_x} \sigma^2(x,1)+\frac{1}{1-\rho_x} \sigma^2(x,0) \Big\}+\Var\{\mu(X_i,1)-\mu(X_i,0)\}.
\end{align*}

If $\{\rho_x, x \in \mathcal{X}\}$ is the solution of optimization (\ref{bound_cnstrt}), then $\hat\tau$ achieves the corresponding bound $v_{\pi^*(\cdot)}$ in Proposition \ref{prop2}.
\end{theorem}
\begin{proof}
Here, we provide a sketch of the proof. For formal details, see Section \ref{secA2} in the Appendix.

Conditional on $X^{(n)}$, Theorem 4 in \citet{ye2024robustness} yields
\begin{equation*}
    \begin{aligned}
        \sqrt{n(x)}\left[\{\hat\mu(x,1)-\hat\mu(x,0)\}-\{\mu(x,1)-\mu(x,0)\}\right]
        \xrightarrow{d}\mathcal N\left(0, \frac{1}{\rho_x} \sigma^2(x,1)+\frac{1}{1-\rho_x} \sigma^2(x,0) \right).
    \end{aligned}
    \end{equation*}
We utilize the decomposition in \citet{Bugni2019} to handle the stratified difference-in-means estimator. Define $\tilde Y_i(w)=Y_i(w)-\mu(X_i,w)$.
    \begin{equation*}
    \begin{aligned}
    \sqrt{n}\left(\hat\tau-\tau\right)= & \sqrt n \Bigg(\sum_{x \in \mathcal X} \frac{n(x)}{n}\Big\{\frac{1}{n(x,1)} \sum_{W_i=1,X_i=x}\tilde{Y}_i(1)-\frac{1}{n(x,0)} \sum_{W_i=0,X_i=x}\tilde{Y}_i(0)\Big\} \\ 
    &+\sum_{i=1}^n \frac{1}{n} [\{\mu(X_i,1)-\mu(X_i,0)\}-\{\mu(1)-\mu(0)\}]\Bigg)\\
    :=&R_{n,1}+R_{n,2},
    \end{aligned}
\end{equation*}
The first term $R_{n,1}$ can be dealt with Theorem 4 in \citet{ye2024robustness}. Conditional on $X^{(n)}$, $n(x)/n\to p(x)$ a.s. and $n(x,1)/n(x)\to \rho_x$ a.s. from Theorem \ref{thm1}. Conditional on $X^{(n)}$, we have in each stratum with $X=x$,

\begin{align*}
    &\sqrt n \frac{n(x)}{n}\Bigg\{\frac{1}{n(x,1)} \sum_{W_i=1,X_i=x}\tilde{Y}_i(1) -\frac{1}{n(x,0)} \sum_{W_i=0,X_i=x}\tilde{Y}_i(0)\Bigg\}\\
    \xrightarrow{d}&\mathcal N \left(0, p(x)\left\{\frac{1}{\rho_x} \sigma^2(x,1)+\frac{1}{1-\rho_x} \sigma^2(x,0)\right\}\right).
\end{align*}
Randomization is independent across strata and hence 
\begin{align*}
    &\sqrt n \sum_{x \in \mathcal X} \frac{n(x)}{n}\Bigg\{\frac{1}{n(x,1)} \sum_{W_i=1,X_i=x}\tilde{Y}_i(1) -\frac{1}{n(x,0)} \sum_{W_i=0,X_i=x}\tilde{Y}_i(0)\Bigg\}\\
    \xrightarrow{d}&\mathcal N \left(0, \sum_{x\in\mathcal X}p(x)\left\{\frac{1}{\rho_x} \sigma^2(x,1)+\frac{1}{1-\rho_x} \sigma^2(x,0)\right\}\right).
\end{align*}

For the second term, $R_{n,2}$ can be dealt with the i.i.d. central limit theorem. We have
\begin{align*}
    \sqrt n \sum_{i=1}^n \frac{1}{n} \left[\{\mu(X_i,1)-\mu(X_i,0)\}-\{\mu(1)-\mu(0)\}\right]
    \xrightarrow{d}\mathcal N(0,\Var\{\mu(X_i,1)-\mu(X_i,0)\}).
\end{align*}

Now we obtain the asymptotic conditional distribution of $R_{n,1}$ given $X^{(n)}$ and the asymptotic distribution of $R_{n,2}$. Because $R^{(n)}$ is the function of $X^{(n)}$, we can obtain the desired result by Lemma \ref{lem:bai} in the Appendix.
\end{proof}
\begin{remark}
    If we restrict our analysis to a single stratum, we arrive at the difference-in-means under the RAR framework, which aligns precisely with the analysis conducted in \citet{ye2024robustness}.
\end{remark}

\begin{remark}
    Aside from stratified DBCD, other CARA designs, such as the CADBCD proposed by \citet{zhang2009new}, may also reach the efficiency bound. In Sections \ref{secC} and \ref{secD2} of the Appendix, we provide both theoretical analysis and empirical simulations of CADBCD, demonstrating its ability to achieve the efficiency bound.
\end{remark}

\section{Numerical Study\label{sec:numerical}}
In this section, we evaluate the empirical performance of the stratified difference-in-means estimator and the difference-in-means estimator under three types of randomization: CR, RAR, and CARA. We consider two cases: without observed covariates and with observed covariates. If covariates are unobserved, we can also use CARA and the stratified difference-in-means, but it will reduce to the difference-in-means under RAR. For RAR and CARA, we select several values of $c$ and report the asymptotic variance bound, total expected outcome $\tilde c$ under adaptive randomization, and the bias and variance of the estimators.

For $w\in\{0,1\}$ and $1\leq i\leq n$, the potential outcomes are generated using the equation:
\[
Y_i(w)=g_w(X_i),
\]
where 
\[g_0(X_i)\sim \begin{cases}
2\times t_5(1), & X_i=1, \\ 
t_5(2)+10, & X_i=2,\\
4\times t_5(3), & X_i=3,
\end{cases}\]
and
\[g_1(X_i)\sim \begin{cases}
t_5(1)+20, & X_i=1, \\ 
3\times t_5(2)+20, & X_i=2,\\
t_5(3)+20, & X_i=3.
\end{cases}\] 
Here, $t_{\nu}(\delta)$ is defined as the distribution of the general non-central $t$ distribution with parameters $(\nu, \delta)$, where $\nu$ is the degree of freedom and $\delta$ is the non-centrality parameter.
We assume $\{X_i\}_{i=1}^n$ are i.i.d., and $g_w(X_i)$ is also i.i.d. 

The conditional treatment effect is \[g_1(X_i)-g_0(X_i)\sim \begin{cases}
20-t_5(1), & X_i=1, \\ 
2\times t_5(2)+10, & X_i=2,\\
20-3\times t_5(3), & X_i=3.
\end{cases}\]
We select this distribution and allow the noncentrality parameter to vary across strata to account for heterogeneity in both variance and mean. Consequently, the allocation probability varies across different strata.

We choose $\gamma=2$ in (\ref{g_func}) to determine the allocation function in the DBCD procedure. The total sample size $n$ is 500. The burn-in number $n_0$ is 10. We replicate the simulation $10^4$ times for each case. The true ATE is numerically calculated based on $10^7$ simulated units, and the bias is defined as the difference between the sample mean of the estimator and the true ATE.  Simulation results for binary response cases, non-optimal allocation rules, and other randomization procedures can be found in Section \ref{secD} of the Appendix.

The results for the case without observed $X$ are shown in Table \ref{tab1}. We use complete randomization (CR) with a 1:1 allocation ratio as a benchmark for comparison. $c=\infty$ means no constraint is imposed. When $c=15$ Neyman allocation still meets the constraint and the bound does not change. When $c\leq14$, Neyman allocation does not meet the constraint and the bound increases as $c$ decreases, indicating stricter constraints lead to higher bounds. Total expected outcome $\tilde c$ meets the constraint $c$ on average for all the scenarios. When $c\leq14$, the constraint is active meaning that the equality holds (e.g., \citealp[p. 128]{boyd2004convex}). In this case, the constraint influences $\tilde c$ by forcing it to be close to $c$. We also find difference-in-means estimator under RAR has the negligible bias and the variance matching the bound. 
\begin{table}[ht]
    \centering
    \caption{Simulation results without observed $X$ (RAR vs. CR).}
    \label{tab1}
    \begin{threeparttable}
    \begin{tabular}{lllcccccccccccccccc}
    \toprule
    & & & &\multicolumn{2}{c}{DIM} \\ \cmidrule(lr){5-6}
    Rand. & c & Bound  & $\tilde{c}$  & \multicolumn{1}{c}{Bias} & \multicolumn{1}{c}{Variance} \\ \midrule
    CR & -- & 0.249 & 16.819 & -0.005 & 0.274 \\
    RAR & $\infty$ & 0.249 & 14.722 & -0.019 & 0.246\\
    RAR & 15 & 0.249 & 14.570 & -0.009 & 0.248\\
    RAR & 14 & 0.252 & 13.959 & 0.008 & 0.256\\
    RAR & 13 &  0.271 & 13.011 & 0.013 & 0.276\\
    \bottomrule    
    \end{tabular}
    \begin{tablenotes}
    \item $c$, constraint of total expected outcome; $\tilde c$, total expected outcome.
    \item Bound, theoretical bound under the constraint of total expected outcome.
    \item DIM, difference-in-means.
    \end{tablenotes}
        
    \end{threeparttable}

\end{table}

Table \ref{tab2} shows the results in the case with observed $X$. We use complete randomization (CR) with a 1:1 allocation ratio as a benchmark for comparison. Similar to Table 1, $c=\infty$ means no constraint is imposed. In this case when $c\leq18$, Neyman allocation does not meet the constraint in all strata, and the bound increases as $c$ decreases, also indicating that stricter constraints lead to higher bounds. In particular, when $c\leq18$ the constraint in stratum with $X=2$ is active and affects $\tilde c_2$. When $c\leq16$ the constraints in strata with $X=2$ and $X=3$ are both active, thus affecting $\tilde c_2$ and $\tilde c_3$. 
In our simulation, the constraint in the stratum with $X=1$ remains inactive. Total expected outcome $\tilde c$ meets the constraint $c$ on average. If and only if the constraint is active, $\tilde c$ is close to $c$. We find that the stratified difference-in-means estimator under CARA has the negligible bias and the variance matching the bound while the difference-in-means estimator under CARA has large bias. 
Note that the large bias of the difference-in-means estimator is expected, as the assignment probabilities differ across strata under CARA. It also implies that achieving the efficiency bound requires both proper design and estimator. Even under constraints, CARA performs better than CR in our considered scenarios. All simulation results support our theoretical results.

\begin{table*}[!htb]
    \centering
  \caption{Simulation results with observed $X$ (CARA vs. CR).}
  \label{tab2}  
    \begin{threeparttable}
    \begin{tabular}{lllcccccccccccccccc}
    \toprule
    & & & & &  &\multicolumn{2}{c}{DIM} &\multicolumn{2}{c}{S-DIM} \\ \cmidrule(lr){7-8}\cmidrule(lr){9-10}
    Rand. & c & Bound  & $\tilde{c}_1$  & $\tilde{c}_2$  & $\tilde{c}_3$  & \multicolumn{1}{c}{Bias} & \multicolumn{1}{c}{Variance} & \multicolumn{1}{c}{Bias} & \multicolumn{1}{c}{Variance}\\ \midrule
    CR & -- & 0.135 & 11.778 &19.746& 18.919& -0.004 & 0.271& -0.003 & 0.170 \\
    %\midrule
    CARA & $\infty$ &  0.135 & 8.573 &23.536 &16.063 & 1.408 & 0.450 &-0.011 & 0.135\\
    CARA & 18& 0.152 & 8.573& 18.017& 16.074 & 0.025 & 0.296& -0.009 & 0.155\\
    CARA &17 & 0.160&  8.573 &17.025& 16.045& -0.291 & 0.288 &-0.005  &0.163\\
    CARA & 16 & 0.174  &8.573 &16.034 &15.852 &-0.675 & 0.289 & 0.004 & 0.182\\
    \bottomrule    
    \end{tabular}
    \begin{tablenotes}
    \item $c$, constraint of total expected outcome; $\tilde c_x$, total expected outcome in stratum with $X=x$.
    \item Bound, variance bound under the constraint of total expected outcome.
    \item DIM, difference-in-means; S-DIM, stratified difference-in-means.
    \end{tablenotes}
        
    \end{threeparttable}

\end{table*}

\section{Conclusion}
Precision medicine customizes healthcare by tailoring treatment to individual patients based on their unique genetic, environmental, and lifestyle information. This approach utilizes biomarkers and adaptive randomization designs to select the most suitable treatments, aiming to optimize preventative and therapeutic care for each person. Due to its incorporation of both responses and covariate information into the randomization process, CARA has garnered increasing attention from both academia and industry.

This paper considers one important type of CARA, specifically implementing DBCD separately within each stratum, and proves that if no additional covariates other than the stratum covariates are observed, the stratified difference-in-means estimator achieves the asymptotic efficiency bound given by \citet{armstrong2022asymptotic}. First, we reviewed the asymptotic efficiency bounds for a broader class of randomization, including CARA, as provided by \citet{armstrong2022asymptotic}. Second, we studied the asymptotic properties of the stratified difference-in-means estimator under the case of implementing DBCD separately within each stratum, utilizing both properties of DBCD and techniques from robust inference \citep{Bugni2019,Bai2022,ye2024robustness}. Building upon these two results, we showed that the stratified difference-in-means estimator with the optimal allocation achieves the asymptotic efficiency bound. Notably, even with the efficient estimator, non-optimal allocation rules could not reach the efficiency bound in general. This finding is supported by the additional simulations in Section \ref{secD2} of the Appendix.

In addition to the case of continuous response, our theoretical results also apply to the binary response case, where the ATE parameter represents the difference between two success probabilities; for more discussions, see Section \ref{secD1} in the Appendix.

It is worth mentioning that we only provide the efficient estimation for one type of CARA, not all CARA designs. Specifically, we require the covariates to be discrete, which may not always be the case in practical clinical trials. Continuous covariates pose challenges for both design and inference properties. In particular, both the conditional target allocation and the conditional mean function are more difficult to estimate. The estimation of the continuous function requires a larger sample size, not to mention the challenges of high-dimensional cases. Few studies focus on robust inference under CARA with continuous covariates, let alone efficient estimation.

In this paper, we bridged the gap between efficiency bounds and efficient estimation for one important type of CARA with discrete covariates. Whether and how to achieve the asymptotic efficiency bound for general CARA with continuous covariates remains an open problem.

\bibliographystyle{apalike}
\bibliography{refs}

\newpage

\appendix

\section{Proof of Main Results\label{secA}}
\subsection{Proof of Proposition \ref{prop2} \label{secA1}}
\begin{proof}
    In this section we define $\pi(x,w)=\pr\{W_i=w|X_i=x\}$ for simplicity of deduction. Notice that optimization problem (\ref{bound_cnstrt}) is equivalent with
    {\small \begin{eqnarray}
\label{bound_cnstrt_armstrong}
\nonumber\mathrm{min}_{\pi(\cdot)}&\ v_{\pi(\cdot)}= \Var\{\mu(X_i,1)-\mu(X_i,0)\}+E\left\{\frac{\sigma^2(X_i,0)}{\pi(X_i,0)}\right\}+E\left\{\frac{\sigma^2(X_i,1)}{\pi(X_i,1)}\right\},\\
\nonumber\text{subject to}&\ E_{\theta^*}\left[r(X_i,1)\pi(X_i,1) + r(X_i,0)\pi(X_i,0)\right]\leq c,\\
\text{and}&\ \pi(x,0)+\pi(x,1)=1 \text{ for all x},
\end{eqnarray}
}
which is restricted and simplified version of (12) in \citet{armstrong2022asymptotic}. Our considered (\ref{bound_cnstrt}) restricts the case to the binary treatment with allocation probabilities adding up to one, as in \citet{hahn2011adaptive}. We follow the proof details in \citet{armstrong2022asymptotic}.

Based on the optimization problem (\ref{bound_cnstrt_armstrong}), we let $\lambda(x)$ and $\eta$ denote the Lagrange multipliers for the constraint. Noting only the second term of $v_{\pi(\cdot)}$ depends on $\pi(\cdot)$, the Lagrangian can be written as
\begin{align*}
\mathcal L=E\Big(\frac{\sigma^2(X_i,0)}{\pi(X_i,0)}+\frac{\sigma^2(X_i,1)}{\pi(X_i,1)}+\eta^\T\left[r(X_i,1)\pi(X_i,1) + r(X_i,0)\pi(X_i,0)\right]+\lambda(X_i)\{\pi(X_i,0)+\pi(X_i,1)-1\}\Big).
\end{align*}
Recall that $\pi^*(x)$ is the solution of this optimization problem (\ref{bound_cnstrt}). Similarly, we define $\pi^*(x,w)$ as the solution of this optimization problem (\ref{bound_cnstrt_armstrong}). Since the equivalence of the two optimization problems, we have $\pi^*(x,1)=\pi^*(x)$ for all $x$. Hence the minimal value of the objective function is also same. We will use $v_{\pi^*(\cdot)}$ interchangeably.

Using the first order condition we have
\begin{equation*}
    \frac{\sigma^2(x,w)}{\pi^*(x,w)^2}=\eta^\T r(x,w) + \lambda(x) \text{ for all $x,w$.}
\end{equation*}
We obtain also
\begin{equation*}
    \pi^*(x,0)+\pi^*(x,1)=1, \text{ for all $x$,}
\end{equation*}
and
\begin{equation*}
    \eta^\T\left[E\{r(X_i,1)\pi^*(X_i,1) + r(X_i,0)\pi^*(X_i,0)\}-c\right]=0.
\end{equation*}
Define score function of $f_X(x)$ and $f_{Y(w)|X}(y|x)$ as $s_X(X_i)$ and $s_w(Y_i(w)|X_i)$, respectively. Let the information for $X_i$ be $I_X=E\{s_X(X_i)s_X(X_i)^\T\}$ and the conditional information for $Y_i(w)$ be $I_{Y(w)|X}=E\{s_w(Y_i(w)|X_i)s_w(Y_i(w)|X_i)^\T|X_i=x\}$.  In the least favorable submodel considered in \citet{Hahn1998} and \citet{armstrong2022asymptotic}, 
\begin{align*}
    I_{Y(w)|X}(x)=&\frac{\sigma^2(x,w)}{\pi^*(x,w)^2}\\
    =&\eta^\T r(x,w) + \lambda(x),
\end{align*}

the corresponding bound is
\begin{align*}
v_{\pi^*(\cdot)}=& I_X+\sum_{w \in \mathcal{W}} E\{\pi^*\left(X_i, w\right) I_{Y(w)| X}(X_i)\} \\
=& I_X+\sum_{w \in \mathcal{W}} E \{\pi^*(X_i, w) \lambda(X_i)\}+\mu^\T \sum_{w \in \mathcal{W}} E \{\pi^*(X_i, w) r(X_i, w)\} \\
=& I_X+E \lambda\left(X_i\right)+\mu^{\T} c,
\end{align*}
which is same as the calculation in page 16 of \citet{armstrong2022asymptotic}. Similarly as the proof of Theorem 5.1 and Corollary 5.1 in \citet{armstrong2022asymptotic}, we can deduce the remaining part.
\end{proof}

\subsection{Proof of Theorem \ref{thm2} \label{secA2}}
\begin{proof}
    Theorem 4 in \citet{ye2024robustness} considered the statistical properties of DBCD and the difference-in-means estimator under DBCD within single stratum---conditional on $n(x)$:
    \begin{equation*}
        n(x,1)/n(x)\to\rho_x,\ \hat{{\theta}}_x\to{\theta}_x,\ \hat\rho_x\to\rho_x\text{\ a.s.},
    \end{equation*}
    \[\hat\mu(x,1)-\hat\mu(x,0)\to \mu(x,1)-\mu(x,0) \text{\ a.s.},\] and 
    \begin{equation*}
    \begin{aligned}
        \sqrt{n(x)}\left[\{\hat\mu(x,1)-\hat\mu(x,0)\}-\{\mu(x,1)-\mu(x,0)\}\right]
        \xrightarrow{d}\mathcal N\left(0, \frac{1}{\rho_x} \sigma^2(x,1)+\frac{1}{1-\rho_x} \sigma^2(x,0) \right).
    \end{aligned}
    \end{equation*}
    From the superpopulation assumption, $n(x)\sim Multinomial (n,p(x))$ for $x\in\mathcal X$, and hence \begin{equation*}
        n(x)/n\to p(x)\text{ a.s.}.
    \end{equation*} 
    Then we have
    \begin{equation*}
        \begin{aligned}
            \hat\tau=\sum_{x\in\mathcal X}\frac{n(x)}{n}\{\hat\mu(x,1)-\hat\mu(x,0)\}
            \to\tau=\sum_{x\in\mathcal X}\frac{n(x)}{n}\{\mu(x,1)-\mu(x,0)\}
            \text{ a.s.}
        \end{aligned}
    \end{equation*}

    In addition, we utilize the decomposition in \citet{Bugni2019} to handle the stratified difference-in-means estimator. Define $\tilde Y_i(w)=Y_i(w)-\mu(X_i,w)$.
    \begin{equation*}
    \begin{aligned}
    \sqrt{n}\left(\hat\tau-\tau\right)= & \sqrt n \Bigg(\sum_{x \in \mathcal X} \frac{n(x)}{n}\Big\{\frac{1}{n(x,1)} \sum_{W_i=1,X_i=x}\tilde{Y}_i(1) -\frac{1}{n(x,0)} \sum_{W_i=0,X_i=x}\tilde{Y}_i(0)\Big\} \\ 
    +&\sum_{i=1}^n \frac{1}{n} \left[\{\mu(X_i,1)-\mu(X_i,0)\}-\{\mu(1)-\mu(0)\}\right]\Bigg)\\
    :=&R_{n,1}+R_{n,2},
    \end{aligned}
\end{equation*}
where the first term can be dealt with Theorem 4 in \citet{ye2024robustness} conditional on $X^{(n)}$ and the second term can be dealt with the i.i.d. central limit theorem. 

From $n(x)/n\to p(x)$ a.s. and $n\to\infty$, for  $x\in\mathcal X$, we have 
\begin{equation*}
\begin{aligned}
    \sup_{u\in\mathbb R}\Bigg|\pr&\Bigg\{\Big\{\frac{1}{\rho_x} \sigma^2(x,1)+\frac{1}{1-\rho_x} \sigma^2(x,0)\Big\}^{-\frac 1 2}\\
    &\times\sqrt {n(x)}\Big\{\frac{1}{n(x,1)} \sum_{W_i=1,X_i=x}\tilde{Y}_i(1) -\frac{1}{n(x,0)} \sum_{W_i=0,X_i=x}\tilde{Y}_i(0)\Big\} \le u \Big| X^{(n)}\Bigg\}\\
    - &\Phi(u)\Bigg|\to0 \text{ a.s.}
\end{aligned}
\end{equation*}
Since $n(x)\in\sigma(X^{(n)})$, we have
\begin{equation*}
\begin{aligned}
    \sup_{u\in\mathbb R}\Bigg|\pr&\Bigg\{\Big[\frac{n(x)}{n}\Big\{\frac{1}{\rho_x} \sigma^2(x,1)+\frac{1}{1-\rho_x} \sigma^2(x,0)\Big\}\Big]^{-\frac 1 2}\\
    &\times\sqrt {n}\frac{n(x)}{n}\Big\{\frac{1}{n(x,1)} \sum_{W_i=1,X_i=x}\tilde{Y}_i(1) -\frac{1}{n(x,0)} \sum_{W_i=0,X_i=x}\tilde{Y}_i(0)\Big\} \le u \Big| X^{(n)}\Bigg\}\\
    - &\Phi(u)\Bigg|\to0 \text{ a.s.}
\end{aligned}
\end{equation*}
Furthermore from Lemma \ref{lem:shao} and $n(x)/n\to p(x)$ a.s., we have
\begin{equation}
\label{eq_within_strara}
\begin{aligned}
    \sup_{u\in\mathbb R}\Bigg|\pr&\Bigg\{\Big[p(x)\Big\{\frac{1}{\rho_x} \sigma^2(x,1)+\frac{1}{1-\rho_x} \sigma^2(x,0)\Big\}\Big]^{-\frac 1 2}\\
    &\times\sqrt {n}\frac{n(x)}{n}\Big\{\frac{1}{n(x,1)} \sum_{W_i=1,X_i=x}\tilde{Y}_i(1) -\frac{1}{n(x,0)} \sum_{W_i=0,X_i=x}\tilde{Y}_i(0)\Big\} \le u \Big| X^{(n)}\Bigg\}\\
    - &\Phi(u)\Bigg|\to0 \text{ a.s.}
\end{aligned}
\end{equation}
From the independence across strata  conditional on $X^{(n)}$ and (\ref{eq_within_strara}), we furthermore have conditional on $X^{(n)}$,
\begin{equation}
\label{eq_r1}
\begin{aligned}
    \sup_{u\in\mathbb R}\Bigg|\pr&\Bigg\{\Big[\sum_{x\in\mathcal X}p(x)\Big\{\frac{1}{\rho_x} \sigma^2(x,1)+\frac{1}{1-\rho_x} \sigma^2(x,0)\Big\}\Big]^{-\frac 1 2}\\
    &\times\sqrt {n}\sum_{x\in\mathcal X}\frac{n(x)}{n}\Big\{\frac{1}{n(x,1)} \sum_{W_i=1,X_i=x}\tilde{Y}_i(1) -\frac{1}{n(x,0)} \sum_{W_i=0,X_i=x}\tilde{Y}_i(0)\Big\} \le u \Big| X^{(n)}\Bigg\}\\
    - &\Phi(u)\Bigg|\to0 \text{ a.s.}
\end{aligned}
\end{equation}
A more understandable way is that conditional on $X^{(n)}$,
\begin{equation*}
\begin{aligned}
    & \sqrt n \sum_{x \in \mathcal X} \frac{n(x)}{n}\Bigg\{\frac{1}{n(x,1)} \sum_{W_i=1,X_i=x}\tilde{Y}_i(1) -\frac{1}{n(x,0)} \sum_{W_i=0,X_i=x}\tilde{Y}_i(0)\Bigg\} \\ 
    \xrightarrow{d} &\mathcal N\left(0,\sum_{x\in\mathcal X}p(x)\left\{\frac{1}{\rho_x} \sigma^2(x,1)+\frac{1}{1-\rho_x} \sigma^2(x,0)\right\}\right).
\end{aligned}
\end{equation*}

For the second term, notice that it is the function of $X^{(n)}$ and we have
\begin{equation}
\begin{aligned}
\label{eq_r2}
    \sqrt n \sum_{i=1}^n \frac{1}{n} \left[\{\mu(X_i,1)-\mu(X_i,0)\}-\{\mu(1)-\mu(0)\}\right]
    \xrightarrow{d}\mathcal N(0,\Var\{\mu(X_i,1)-\mu(X_i,0)\}).
\end{aligned}
\end{equation}
From (\ref{eq_r1}), (\ref{eq_r2}) and Lemma \ref{lem:bai}, we obtain the result
\begin{align*}
    \sqrt{n}\left(\hat\tau-\tau\right)
    \xrightarrow{d}\mathcal N\Bigg(0,\sum_{x\in\mathcal X}p(x)\left\{\frac{1}{\rho_x} \sigma^2(x,1)+\frac{1}{1-\rho_x} \sigma^2(x,0) \right\}+\Var\{\mu(X_i,1)-\mu(X_i,0)\}\Bigg).
\end{align*}

\end{proof}

\section{Auxiliary Lemmas\label{secB}}
\begin{lemma}
\label{lem:shao}
    For $n\geq 1$, let $Y_n$ be some multivaraite random variable, $B_n=B_n(Y_n)\in\sigma(Y_n)$ and $Z_n$ be two random variables, and $\Phi(\cdot)$ be the standard normal c.d.f. Suppose that \[
\sup_{u\in\mathbb R}\mid P\{Z_n/B_n\leq u|Y_n\} - \Phi(u)|\to 0
\] holds a.s. and  additionally  $B_n\to B>0$ a.s.
    Then we have
    \[
\sup_{u\in\mathbb R}\mid P\{Z_n/B\leq u|Y_n\} - \Phi(u)|\to 0 \text{\ a.s.}
\]
\end{lemma}
\begin{proof}
    We prove the almost sure convergence.
    
    Denote 
    \begin{align*}
        \Omega_1
        =\{\omega:\sup_{u\in\mathbb R}| P\{Z_n/B_n\leq u|Y_n(\omega)\}- \Phi(u)|\to 0\}
    \end{align*}
    and \[\Omega_2=\{\omega: B_n(w)=B_n(Y_n(\omega))\to B \}.\] By the almost sure definition we have $P(\Omega_1\cap\Omega_2)=1$.

We have
\[
\sup_{u\in\mathbb R}\mid P\{Z_n/B_n(Y_n(\omega))\leq u|Y_n(\omega)\} - \Phi(u)|\to 0,
\]
for any $\omega\in\Omega_1$. Since $\omega$ is fixed and $B_n(\omega)=B_n(Y_n(\omega))$ is also fixed, equivalently, we have
\[
\sup_{u\in\mathbb R}\mid P\{Z_n\leq u|Y_n(\omega)\} - \Phi(u/B_n(\omega))|\to 0,
\]
for any $\omega\in\Omega_1$. 

Furthermore we claim that
\[
\sup_{u\in\mathbb R}\mid \Phi(u/B) - \Phi(u/B_n(\omega))|\to 0
\]
for any $\omega\in\Omega_2$.

Hence for any $\omega\in\Omega_1\cap\Omega_2$
\begin{align*}
&\sup_{u\in\mathbb R}\mid P\{Z_n\geq u|Y_n(\omega)\} - \Phi(u/B) \mid
\\
\leq&\sup_{u\in\mathbb R}\mid P\{Z_n\geq u|Y_n(\omega)\} - \Phi(u/B_n(\omega))|+ \sup_{u\in\mathbb R}\mid \Phi(u/B) - \Phi(u/B_n(\omega)) \mid\\
\to&0.
\end{align*}
Since $P(\Omega_1\cap\Omega_2)=1$ we get the almost sure convergence.

To prove the claim, we define a series of i.i.d. $W_n\sim \mathcal N(0,1)$ which are independent with the probability space $(\Omega,\mathcal F,P)$.
For any fixed $\omega\in\Omega_2$, $B_n(\omega)\to B>0$ and then by Slutsky Theorem we obtain that
\[
B_n(\omega)W_n \to \mathcal N(0,B^2),
\]
in distribution
and equivalently,
\[
\sup_{u\in\mathbb R}\mid \Phi(u/B) - \Phi(u/B_n(\omega))|\to 0.
\]
\end{proof}

We also give one similar but more general lemma than Lemma S.1.2 in \citet{Bai2022}.
\begin{lemma}
\label{lem:bai}
    For $n\geq 1$, let $U_n$ and $V_n$ be real-valued random variables and $\mathcal F_n$ a $\sigma-$field generated by multivariate random variable $Y_n$. Suppose
    \[
    P\{U_n\leq u\mid \mathcal F_n\}\to \Phi(u/\tau_1)
    \]
    in probability, where $\Phi(\cdot)$ is the standard normal c.d.f. and $\tau_1,\tau_2$ are constants. Further assume $V_n$ is $\mathcal F_n$- measurable and 
    \[
    P\{V_n\leq u\}\to \Phi(u/\tau_2).
    \]
    Then
    \[
    P\{U_n+V_n\leq u\}\to\Phi(u/\sqrt{\tau_1^2+\tau_2^2}).
    \]
\end{lemma}
\begin{proof}
From the convergence in probability, we choose one sub-array $\{n_k\}_{k=1}^\infty$ converging almost surely. Then without loss of generality, 
    we prove the a.s. convergence.

    Denote $\Omega_1=\{\omega:P\{U_n\leq u\mid \mathcal F_n\}\to \Phi(u/\tau_1/2)\}$, where $P(\Omega_1)=1$.
    In the conditional probability space $\{Y_n=Y_n(w)\}$ where $\omega\in\Omega_1$, characteristic function converges too. Hence
    \[
    E\{\exp(itU_n)\mid Y_n\}\to \exp(it\tau_1^2/2) \ \mathrm{a.s.}
    \]
    $V_n$ converges to $\mathcal N(0,\tau_2^2)$. Then we have
    \[
    E\{\exp(itV_n)\}\to \exp(it\tau_2^2/2).
    \]
    Hence
    \begin{align*}
        E[\exp\{it(V_n+U_n)\}]=&E(E[\exp\{it(V_n+U_n)\}\mid \mathcal F_n])\\
        =&E[E\{\exp(itU_n)\mid \mathcal F_n\}\exp(itV_n)]\\
        \to&\exp\{it(\tau_1^2+\tau_2^2)/2\}.
    \end{align*}
    The three lines come from the tower property, $V_n\in\mathcal F_n$ and the dominated convergence theorem, respectively.

\end{proof}

\section{Stratified Difference-in-means Estimator under CADBCD\label{secC}}
\begin{theorem}
    Under the CADBCD design \citep{zhang2009new}, if Assumptions 1--4 are satisfied, the stratified difference-in-means estimator with the optimal allocation achieves the efficiency bound in \citet{armstrong2022asymptotic}.
\end{theorem}
\begin{proof}
Under the CADBCD design, the allocation probability for the $(m+1)$-th subject with covariate $X_{m+1}$ is
\[
\frac{\hat{\rho}_{X_{m+1},m}\left(\frac{\sum_{k=1}^m\hat{\rho}_{X_{k},m}}{N_{m, 1}}\right)^\gamma}{\hat{\rho}_{X_{m+1},m}\left(\frac{\sum_{k=1}^m\hat{\rho}_{X_{k},m}}{N_{m, 1} }\right)^\gamma+(1-\hat{\rho}_{X_{m+1},m})\left(\frac{m-\sum_{k=1}^m\hat{\rho}_{X_{k},m}}{m-N_{m, 1}}\right)^\gamma},
\]
where $\gamma\ge0$ is the parameter reflecting the degree of randomness, $N_{m,1}$ is the number of first $m$ subjects assigned to $W=1$ and $\hat\rho_{x,m}$ is the estimate of $\rho_x$ based on the first $m$ subjects.

Under Assumptions 1--4, and following a similar approach to the proof of Theorem 1, it follows from the properties of the CADBCD design that 
\[\hat{{\theta}}_x\to{\theta}_x,\ \hat\rho_x\to\rho_x \text{ a.s.}\] 
for all $x\in\mathcal X$.
Furthermore, the allocation probabilities with the covariate $x$ converges to $\rho_{x}$ a.s. Hence $n(x,1)/n(x)\to\rho_x$ a.s.
    
We utilize the following decomposition to handle the stratified difference-in-means estimator. Define $\tilde Y_i(w)=Y_i(w)-\mu(X_i,w)$.
    \begin{align*}
    \sqrt{n}\left(\hat\tau-\tau\right)= & \sqrt n \Bigg(\sum_{x \in \mathcal X} \frac{n(x)}{n}\Big\{\frac{1}{n(x,1)} \sum_{W_i=1,X_i=x}\tilde{Y}_i(1) %\\&\quad\quad\quad\quad\quad\quad\quad
    -\frac{1}{n(x,0)} \sum_{W_i=0,X_i=x}\tilde{Y}_i(0)\Big\} \notag\\ 
    &+\sum_{i=1}^n \frac{1}{n} [\{\mu(X_i,1)-\mu(X_i,0)\}%\\ &\quad\quad\quad\quad\quad\quad\quad
    -\{\mu(1)-\mu(0)\}]\Bigg)\notag\\
    =&\frac{1}{\sqrt{n}}\sum_{i=1}^n\Bigg(\left\{ \frac{1}{\rho_{X_i}}W_i \tilde Y_i(1)-\frac{1}{1-\rho_{X_i}}(1-W_i)\tilde Y_i(0)\right\}\notag\\
&\quad\quad+\Bigg[\left\{\frac{n(X_i)}{n(X_i,1)}-\frac{1}{\rho_{X_i}}\right\}W_i \tilde Y_i(1)%\\&\quad\quad\quad\quad
-\left\{\frac{n(X_i)}{n(X_i,0)}-\frac{1}{1-\rho_{X_i}}\right\}(1-W_i) \tilde Y_i(0)\Bigg]\notag\\
&\quad\quad+[\{\mu(X_i,1)-\mu(X_i,0)\}-\{\mu(1)-\mu(0)\}]\Bigg)\notag\\
    :=&R_{n,1}+R_{n,2}+R_{n,3}.
    \end{align*}

First we show that $R_{n,2}$ is $o_p(1)$.  we have
\begin{align*}
    &\frac{1}{\sqrt n}\sum_{i=1}^n\left\{\frac{n(X_i)}{n(X_i,1)}-\frac{1}{\rho_{X_i}}\right\}W_i \tilde Y_i(1)\\
    =&\frac{1}{\sqrt n}\sum_{x\in\mathcal X}\left\{\frac{n(x)}{n(x,1)}-\frac{1}{\rho_{x}}\right\}\sum_{X_i=x}W_i \tilde Y_i(1).
\end{align*}
For $x\in\mathcal X$, define $n_k(x,w)$ as the number of units for the first $k$ units in stratum $x$ and treatment $w$. Define $\tau_j{(x,w)}=\text{min}\{k:n_k(x,w)=j\}$ where $\min\{\emptyset\}=+\infty$. With the argument of \citet{doob1936note}, we can construct one i.i.d. sequence $\check Y_i(1)$ which coincides with $\tilde Y_i(1)$ on the event $\{n(x,1)\to\infty \text{ for all } x\in\mathcal X\}$. Let $\check Y_i(x,w)= \{Y_{\tau_i(x,w)}(w)-\mu(x,w)\}I\{\tau_i(x,w)<\infty\}+\xi_i(x,w)I\{\tau_i(x,w)=\infty\})$,  where $\{\xi_i(x,w)\}$ is 
 an independent copy of $\{Y_i(w)-\mu(x,w)\}$, which is also independent of $\{W_i\}$. Then $\check Y_i(x,w)$ is a sequence of i.i.d. random variables with the same distribution as $\tilde Y_1(w)|X_1=x$.
 
 Since for all $x\in\mathcal X$, $\frac{n(x,1)}{n(x)}\to\rho_{x}$ and $n(x)\to\infty$ a.s., we have
\begin{align*}
    \frac{1}{\sqrt{n}}\sum_{X_i=x}W_i\tilde Y_i(1)=&\frac{1}{\sqrt{n}}\sum_{1\le j\le n(x,1)}\check Y_j(x,1)\\
    \xrightarrow{d}&\mathcal N(0,p_x\sigma^2(x,1))\\
    =&O_P(1).
\end{align*}
Then we  have
\begin{align*}
    \frac{1}{\sqrt n}\left\{\frac{n(x)}{n(x,1)}-\frac{1}{\rho_{x}}\right\}\sum_{X_i=x}W_i \tilde Y_i(1)=&o_P(1)\cdot O_P(1)\\
    =&o_P(1).
\end{align*}
Since $x$ is discrete, we combine the terms for each $x\in\mathcal X$.
\begin{align*}
    &\frac{1}{\sqrt n}\sum_{x\in\mathcal X}\left\{\frac{n(x)}{n(x,1)}-\frac{1}{\rho_{x}}\right\}\sum_{X_i=x}W_i \tilde Y_i(1)=o_P(1).
\end{align*}
Similarly we have
\begin{align*}
    &\frac{1}{\sqrt n}\sum_{x\in\mathcal X}\left\{\frac{n(x)}{n(x,0)}-\frac{1}{1-\rho_{x}}\right\}\sum_{X_i=x}(1-W_i) \tilde Y_i(0)=o_P(1).
\end{align*}
We obtain that $R_{n,2}=o_P(1)$.

Note that $R_{n,3}$ is function of $X^{(n)}$ and has the asymptotic normality.
\begin{align*}
    &\sqrt n \sum_{i=1}^n \frac{1}{n} \left[\{\mu(X_i,1)-\mu(X_i,0)\}-\{\mu(1)-\mu(0)\}\right]\\
    \xrightarrow{d}&\mathcal \mathcal N(0,\Var\{\mu(X_i,1)-\mu(X_i,0))\}.
\end{align*}
From the similar idea as the proof of Theorem 2, it is sufficient to show that $R_{n,1}|X^{(n)}$ conditionally converges  to $\mathcal N\left(0,\sum_{x\in\mathcal X}p(x)\left\{\frac{1}{\rho_x} \sigma^2(x,1)+\frac{1}{1-\rho_x} \sigma^2(x,0)\right\}\right)$.

Given $X^{(n)}$, define 
$\mathcal F_i=\sigma(W_1,\cdots,W_i;Y_1,\cdots,Y_i)$ and $\mathcal G_i=\sigma(W_1,\cdots,W_i;Y_1,\cdots,Y_i;X_1,\cdots,X_i,X_{i+1})$. We have
\begin{align*}
    R_{n,1}=&\frac{1}{\sqrt n}\sum_{i=1}^n \{\frac{W_i}{ \rho_{X_i}} \tilde Y_i(1)-\frac{1-W_i}{1-\rho_{X_i}}\tilde Y_i(0)\}\\
    :=&\frac{1}{\sqrt n}\sum_i (\Delta Q_{i(1)}-\Delta Q_{i(0)}).
\end{align*}
Since the $X_i$ is i.i.d. and $W_i$ is only dependent on $\mathcal G_i$, we have that given $X^{(n)}$, $\{\Delta Q_{i(w)},\mathcal F_{i-1}; i=1,2,\cdots,n\}$ is martingale difference for $w\in\mathcal{W}$. 
\begin{align*}
    &\Var\{\Delta Q_{i(1)}|\sigma(\mathcal F_{i-1}, X^{(n)})\}\\
    =&E\{\frac{W_i}{\rho_{X_i}^2} \tilde Y^2_i(1)|\sigma(\mathcal F_{i-1}, X^{(n)})\}\\
    =&\frac{\sigma^2(X_i,1)}{\rho_{X_i}^2}E\{W_i |\sigma(\mathcal F_{i-1}, X^{(n)})\}\\
    =&\frac{\sigma^2(X_i,1)}{\rho_{X_i}^2}E\{W_i |\mathcal G_{i-1}\} \\
    =&\frac{\sigma^2(X_i,1)}{\rho_{X_i}^2}\rho_{X_i} + o(1) \text{ a.s. }\\
    =&\frac{\sigma^2(X_i,1)}{\rho_{X_i}}+ o(1) \text{ a.s. }
\end{align*}
Similarly we have 
\[
\Var\{\Delta Q_{i(0)}|\sigma(\mathcal F_{i-1}, X^{(n)})\}=\frac{\sigma^2(X_i,0)}{1-\rho_{X_i}}\ + o(1) \text{ a.s. }
\]
Notice that
\begin{align*}
    &\Cov\{\Delta Q_{i(1)},\Delta Q_{i(0)}|\sigma(\mathcal F_{i-1}, X^{(n)})\}\\
    =&E\{\frac{W_i}{\rho_{X_i}} \tilde Y_i(1) \frac{1-W_i}{1-\rho_{X_i}} \tilde Y_i(0)|\sigma(\mathcal F_{i-1}, X^{(n)})\}\\
    =&0.
\end{align*}
Then we have
\begin{align*}
&\frac{1}{n}\sum_{i=1}^n \Var\{\Delta Q_{i(1)}-\Delta Q_{i(0)}|\sigma(\mathcal F_{i-1}, X^{(n)})\}\\
=&\sum_i \{\frac{1}{\rho_{X_i}}\sigma^2(X_i,1)+\frac{1}{1-\rho_{X_i}}\sigma^2(X_i,0)\}\\
\to&\sum_{x\in\mathcal X}p(x)\left\{\frac{1}{\rho_x} \sigma^2(x,1)+\frac{1}{1-\rho_x} \sigma^2(x,0)\right\}\text{ a.s.}
\end{align*}

Note that given $X^{(n)}$, $n(x,1)/n(x)\to \rho_x$ a.s. uniformly for all $x\in\mathcal X$. On the other hand, for some $\v>0$,
\begin{align*}
    &\frac{1}{n^{1+\v/2}}\sum_i E \{|\Delta Q_{i(1)}-\Delta Q_{i(0)}|^{2+\v}|\sigma(\mathcal F_{i-1}, X^{(n)})\}\\
    \leq &\frac{1}{n^{1+\v/2}}\sum_i E \{|\Delta Q_{i(1)}|^{2+\v}+|\Delta Q_{i(0)}|^{2+\v}|\sigma(\mathcal F_{i-1}, X^{(n)})\}\\
    \leq&\frac{1}{n^{1+\v/2}}[\sum_i \{\frac{E\{\tilde Y_i(1)|^{2+\v}|X_i\}}{\rho_{X_i}^{2+\v}},+\frac{E\{\tilde Y_i(0)|^{2+\v}|X_i\}}{(1-\rho_{X_i})^{2+\v}}\}+o(1)] \text{ a.s.}\\
    \leq & \frac{C_1 + o(1) }{n^{\v/2}} \text{ a.s.} \\
    \to&0 \text{ a.s.}
\end{align*}
where $C_1$ is some positive constant. Hence the conditional Lyapunov condition holds. Therefore by the central limit theorem of the martingale, we have that conditional on $X^{(n)}$,
\[
R_{n,1}\xrightarrow{d}\mathcal N\left(0,\sum_{x\in\mathcal X}p(x)\left\{\frac{1}{\rho_x} \sigma^2(x,1)+\frac{1}{1-\rho_x} \sigma^2(x,0)\right\}\right).
\]
Finally we use Lemma 2 to combine $R_{n,1}$ and $R_{n,2}$.

\end{proof}

\section{Additional Simulations\label{secD}}

\subsection{Additional simulation results with the binary response\label{secD1}}
In this subsection, we consider two strata, where heterogeneity of the treatment effect exists between them.
More specifically, for $w\in\{0,1\}$ and $1\leq i\leq n$, the potential outcomes are generated using the equation:
\[
Y_i(w)=g_w(X_i),
\]
where 
\[g_0(X_i)\sim\text{Bernoulli}(0.5)\]
and
\[g_1(X_i)\sim \begin{cases}
\text{Bernoulli}(0.9), & X_i=1, \\ 
\text{Bernoulli}(0.5), & X_i=2.
\end{cases}\] 
The other settings are the same as those in the main text. We use complete randomization (CR) with a 1:1 allocation ratio as the baseline for comparison. Similar to the case of continuous response, the simulation results with binary response also support our theoretical results.

The results for the case without observed $X$ are shown in Table \ref{tab3}. When $c\leq0.60$, Neyman allocation does not meet the constraint and the bound increases as $c$ decreases, indicating stricter constraints lead to higher bounds. In other words, the constraint is active in sense of \citet{boyd2004convex}, i.e., affects $\tilde c$ and forced it close to $c$. Total expected outcome $\tilde c$ meets the constraint $c$ on average for all the scenarios. We also find difference-in-means estimator under RAR has the negligible bias and the variance matching the bound.

\begin{table}[ht]
    \centering
    \caption{Simulation results with binary response and no observed $X$ (RAR vs. CR).}
    \label{tab3}
    \begin{threeparttable}
    \begin{tabular}{lllcccccccccccccccc}
    \toprule
    & & & &\multicolumn{2}{c}{DIM} \\ \cmidrule(lr){5-6}
    Rand. & c & Bound  & $\tilde{c}$  & \multicolumn{1}{c}{Bias} & \multicolumn{1}{c}{Variance} \\ \midrule
    CR & -- & 0.00170&  0.63352& -0.00040&  0.00173 \\
    RAR & $\infty$ & 0.00170& 0.62237& 0.00045& 0.00170\\
    RAR & 0.65 & 0.00170& 0.62120& 0.00093& 0.00168\\
    RAR & 0.60 & 0.00175& 0.59912& 0.00538& 0.00173\\
    RAR & 0.55 &  0.00253 &0.56704 &0.00466 &0.00241\\
    \bottomrule    
    \end{tabular}
    \begin{tablenotes}
    \item $c$, constraint of total expected outcome; $\tilde c$, total expected outcome.
    \item Bound, theoretical bound under the constraint of total expected outcome.
    \item DIM, difference-in-means.
    \end{tablenotes}
        
    \end{threeparttable}

\end{table}

Table \ref{tab4} shows the results in the case with observed $X$. In this case, when $c\leq0.60$ the constraint in stratum with $X=1$ is active and affects $\tilde c_1$.
In our simulation, the constraint in the stratum with $X=2$ remains inactive. Total expected outcome $\tilde c$ meets the constraint $c$ on average.  We find that the stratified difference-in-means estimator under CARA has the negligible bias and the variance matching the bound while the difference-in-means estimator under CARA has large bias.

\begin{table*}[!htb]
    \centering
  \caption{Simulation results with binary response and observed $X$ (CARA vs. CR).}
  \label{tab4}  
    \begin{threeparttable}
    \begin{tabular}{lllcccccccccccccccc}
    \toprule
    & & & & & &\multicolumn{2}{c}{DIM} &\multicolumn{2}{c}{S-DIM} \\ \cmidrule(lr){7-8}\cmidrule(lr){9-10}
    Rand. & c & Bound  & $\tilde{c}_1$  & $\tilde{c}_2$  & \multicolumn{1}{c}{Bias} & \multicolumn{1}{c}{Variance} & \multicolumn{1}{c}{Bias} & \multicolumn{1}{c}{Variance}\\ \midrule
    CR & -- & 0.00159& 0.70012&0.49938 &0.00066& 0.00173& 0.00055& 0.00167 \\
    %\midrule
    CARA & $\infty$ & 0.00159&  0.64510  &0.50039 &-0.02991 & 0.00158 & 0.00144 & 0.00163\\
    CARA & 0.65& 0.00159 & 0.63652 & 0.50042& -0.03566 & 0.00202 &0.00219 & 0.00160\\
    CARA & 0.60 & 0.00166 & 0.60183 & 0.50037 &-0.06644&  0.00315 & 0.00265 & 0.00168\\
    CARA & 0.55 & 0.00209 & 0.57711 & 0.50033& -0.10026 & 0.00301 &-0.00034  &0.00204\\
    \bottomrule    
    \end{tabular}
    \begin{tablenotes}
    \item $c$, constraint of total expected outcome; $\tilde c_x$, total expected outcome in stratum with $X=x$.
    \item Bound, variance bound under the constraint of total expected outcome.
    \item DIM, difference-in-means; S-DIM, stratified difference-in-means.
    \end{tablenotes}
        
    \end{threeparttable}

\end{table*}

\subsection{Comparisons with non-optimal allocations and other randomization methods\label{secD2}}

In this subsection we focus on the comparisons with non-optimal allocations and other randomization methods. We use the same model and setting as those in the main text.

In the case with no constraint and observed $X$, we consider complete randomization (1:1 allocation ratio), Efron's biased coin design \citep{Efron1971} (1:1 allocation ratio and 0.75 biased coin probability), RAR with optimal allocation, and two additional non-optimal allocations, one proposed by \citet{bandyopadhyay2001adaptive} and the RSIHR allocation \citep[p. 13]{rosenberger2001,Hu2006}, respectively.
Assume that the response is normal and $\sigma_w^2$ and $\mu_w$ are the variance and expectation for treatment $w, w\in\{0,1\}$, respectively, then the target allocation proposed by \citet{bandyopadhyay2001adaptive} has the following form:

$$
\rho({\theta})=\Phi\left(\frac{\mu_1-\mu_0}{T}\right), {\theta}=\left(\mu_1, \mu_0\right),
$$
where $\Phi(\cdot)$ is the cumulative density function of standard normal distribution and $T$ is a constant set by 30 in the simulation.
The generalized RSIHR allocation \citep[p. 13]{Hu2006} is in the form
$$
\rho({\theta})=
\frac{\sigma_1 \sqrt{\mu_0}}{\sigma_1 \sqrt{\mu_0}+\sigma_0 \sqrt{\mu_1}}
$$
where ${\theta}=\left(\mu_1, \sigma_1^2, \mu_0, \sigma_0^2\right)$. When the response is binary, the generalized RSIHR allocation will reduce to the original and familiar form \citep{rosenberger2001}
$$
\rho({\theta})=
\frac{ \sqrt{\mu_0}}{ \sqrt{\mu_0}+ \sqrt{\mu_1}}
$$
where ${\theta}=\left(\mu_1, \mu_0\right)$ since $\sigma_1=\sigma_0=\sqrt{\mu_1\mu_0}$ in the binary case.

\begin{table}[ht]
    \centering
    \caption{Comparisons with non-optimal allocations and
other randomization methods (without observed $X$).}
    \label{tab5}
    \begin{threeparttable}
    \begin{tabular}{lllcccccccccccccccc}
    \toprule
    & & & &\multicolumn{2}{c}{DIM} \\ \cmidrule(lr){5-6}
    Rand. & c & Bound  & $\tilde{c}$  & \multicolumn{1}{c}{Bias} & \multicolumn{1}{c}{Variance} \\ \midrule
    CR & -- & 0.249& 16.819 &-0.005&  0.274 \\
    BCD & -- & 0.249& 16.820 &-0.006 & 0.267\\
    $\text{RAR}_{\text{Optimal}}$ & -- & 0.249& 14.722& -0.019 & \textbf{0.246}\\
    $\text{RAR}_\text{BandBis}$ & -- & 0.249& 19.435 & 0.003 & 0.367\\
    $\text{RAR}_\text{RSIHR}$ & -- & 0.249& 13.324 &-0.028 & 0.267\\
    \bottomrule    
    \end{tabular}
    \begin{tablenotes}
    \item $c$, constraint of total expected outcome; $\tilde c$, total expected outcome.
    \item Bound, theoretical bound under the constraint of total expected outcome.
    \item DIM, difference-in-means.
    \item BCD, Efron's biased coin design.
    \item $\text{RAR}_\text{*}$, RAR based on the allocation rule $*$.
    \end{tablenotes}
        
    \end{threeparttable}

\end{table}

In the case with no constraint but observed $X$, we consider complete randomization (1:1 allocation ratio), minimization method \citep{Pocock1975} (1:1 allocation ratio, with equal weight and 0.75 biased coin probability), stratified DBCD and CADBCD \citep{zhang2009new} with optimal allocation and other two non-optimal allocations, one proposed by \citet{bandyopadhyay2001adaptive} and the RSIHR allocation \citep[p. 13]{rosenberger2001,Hu2006}, respectively. We let $\gamma=2$ in stratified DBCD and CADBCD.

The results are summarized in Tables \ref{tab5} and \ref{tab6}, with the optimal variances highlighted in bold. Both minimization and Efron’s biased coin design slightly outperform complete randomization but still perform worse than CARA and RAR with the optimal allocation. 
The simulations demonstrate that our proposed method, stratified DBCD with optimal allocation, achieves the lowest variance across all methods. Both RAR and CARA with optimal allocation outperform their counterparts using non-optimal allocation rules in every scenario. The simulations suggest that non-optimal methods do not achieve the efficiency bound, nor do CAR methods like Efron's biased coin design and minimization in general.

\begin{table*}[!htb]
    \centering
  \caption{Comparisons with non-optimal allocations and
other randomization methods (with observed $X$, 500 units).}
  \label{tab6}  
    \begin{threeparttable}
    \begin{tabular}{lllcccccccccccccccc}
    \toprule
    & & & & &  &\multicolumn{2}{c}{DIM} &\multicolumn{2}{c}{S-DIM} \\ \cmidrule(lr){7-8}\cmidrule(lr){9-10}
    Rand. & c & Bound  & $\tilde{c}_1$  & $\tilde{c}_2$  & $\tilde{c}_3$  & \multicolumn{1}{c}{Bias} & \multicolumn{1}{c}{Variance} & \multicolumn{1}{c}{Bias} & \multicolumn{1}{c}{Variance}\\ \midrule
    CR & -- & 0.135 &11.778 &19.746 &18.919 &-0.004 & 0.271& -0.003 & 0.170 \\
    %\midrule
    MIN & -- & 0.135 &11.787& 19.757& 18.919 &-0.006 & 0.176 &-0.005 & 0.175\\
    $\text{SDBCD}_\text{Optimal}$ & --& 0.135  &8.573& 23.536& 16.063 & 1.408 & 0.450& -0.011 & \textbf{0.135}\\
    $\text{SDBCD}_\text{BandBis}$ & -- & 0.135& 16.168 &22.519& 20.044 &-0.714 & 0.294 & 0.005 & 0.191\\
    $\text{SDBCD}_\text{RSIHR}$ & -- & 0.135 & 5.057& 22.330 &15.737  &2.463  &0.395 &-0.023&  0.141\\
    $\text{CADBCD}_{\text{Optimal}}$ & --& 0.135 & 8.773& 22.861& 16.378 & 1.233 & 0.498& -0.008 & 0.140\\
    $\text{CADBCD}_{\text{BandBis}}$ &-- & 0.135& 15.993 &22.492 &20.123 &-0.592&  0.410 & 0.011  &0.196\\
    $\text{CADBCD}_{\text{RSIHR}}$ & -- & 0.135 & 5.675 &21.367 &16.020 & 2.009  &0.429 &-0.014 & 0.143 \\

    \bottomrule    
    \end{tabular}
    \begin{tablenotes}
    \item $c$, constraint of total expected outcome; $\tilde c_x$, total expected outcome in stratum with $X=x$.
    \item Bound, variance bound under the constraint of total expected outcome.
    \item DIM, difference-in-means; S-DIM, stratified difference-in-means.
    \item MIN, minimization method.
    \item $\text{SDBCD}_\text{*}$, stratified DBCD based on the allocation rule $*$.
    \item $\text{CADBCD}_\text{*}$, CADBCD based on the allocation rule $*$.
    \end{tablenotes}
        
    \end{threeparttable}

\end{table*}

Notably, in our simulation, CADBCD performs comparably to stratified DBCD across all three allocation rules. We also present a larger-sample simulation with 2000 units, compared to the original 500 units; see Table \ref{tab7}. The results show that both stratified DBCD and CADBCD with the optimal allocation achieve the efficiency bound, while other randomization methods do not, as expected. The simulation coincides with our theoretical results.

\begin{table*}[!htb]
    \centering
  \caption{Comparisons with non-optimal allocations and
other randomization methods (with observed $X$).}
  \label{tab7}  
    \begin{threeparttable}
    \begin{tabular}{lllcccccccccccccccc}
    \toprule
    & & & & &  &\multicolumn{2}{c}{DIM} &\multicolumn{2}{c}{S-DIM} \\ \cmidrule(lr){7-8}\cmidrule(lr){9-10}
    Rand. & c & Bound  & $\tilde{c}_1$  & $\tilde{c}_2$  & $\tilde{c}_3$  & \multicolumn{1}{c}{Bias} & \multicolumn{1}{c}{Variance} & \multicolumn{1}{c}{Bias} & \multicolumn{1}{c}{Variance}\\ \midrule
    CR & -- & 0.0339& 11.7874& 19.7567& 18.9202& -0.0075&  0.0678& -0.0064 & 0.0418 \\
    %\midrule
    MIN & -- & 0.0339 &11.7859 &19.7568& 18.9208& -0.0079  &0.0421& -0.0078  &0.0421\\
    $\text{SDBCD}_\text{Optimal}$ & --& 0.0339&  8.6210& 23.4984 &16.0960&  1.3810 & 0.1217 &-0.0080  &\textbf{0.0336}\\
    $\text{SDBCD}_\text{BandBis}$ & -- & 0.0339 &16.1928& 22.5410 &20.0524& -0.7244&  0.0764 &-0.0024 & 0.0493\\
    $\text{SDBCD}_\text{RSIHR}$ & -- & 0.0339  &5.0282& 22.2962 &15.7484 & 2.4695&  0.1104 &-0.0121 & 0.0351\\
    $\text{CADBCD}_{\text{Optimal}}$ & --& 0.0339  &8.6669 &23.3314 &16.1721&  1.3383&  0.1479& -0.0090 & \textbf{0.0338}\\
    $\text{CADBCD}_{\text{BandBis}}$ &-- & 0.0339 &16.1521 &22.5299& 20.0741 &-0.6987 & 0.1041& -0.0046 & 0.0489\\
    $\text{CADBCD}_{\text{RSIHR}}$ & -- & 0.0339  &5.2004& 22.0493 &15.8204 & 2.3461 & 0.1244 &-0.0122 & 0.0352 \\

    \bottomrule    
    \end{tabular}
    
    \begin{tablenotes}
    \item $c$, constraint of total expected outcome; $\tilde c_x$, total expected outcome in stratum with $X=x$.
    \item Bound, variance bound under the constraint of total expected outcome.
    \item DIM, difference-in-means; S-DIM, stratified difference-in-means.
    \item MIN, minimization method.
    \item $\text{SDBCD}_\text{*}$, stratified DBCD based on the allocation rule $*$.
    \item $\text{CADBCD}_\text{*}$, CADBCD based on the allocation rule $*$.
    \end{tablenotes}
        
    \end{threeparttable}

\end{table*}

\subsection{Estimated allocation probabilities based on burn-in samples and all samples\label{secD3}}
In this subsection, we show the estimated allocation probabilities affected by the different constraint levels and compare the allocation probabilities estimated with burn-in samples and all samples. We use the same model and setting as those in the main text with 2000 replications.

Refer to Figure \ref{Fig::rar} for the results under RAR. The true allocation probability decreases as the constraint level tightens. In the simulation, both the all-sample estimates and the burn-in estimates are generally close to the true values, though the all-sample estimates show lower variances. This suggests that a burn-in size of 10 is effective for start-up design.

\begin{figure*}[htb]
  \begin{center}
    \includegraphics[width=150mm,height=90mm]{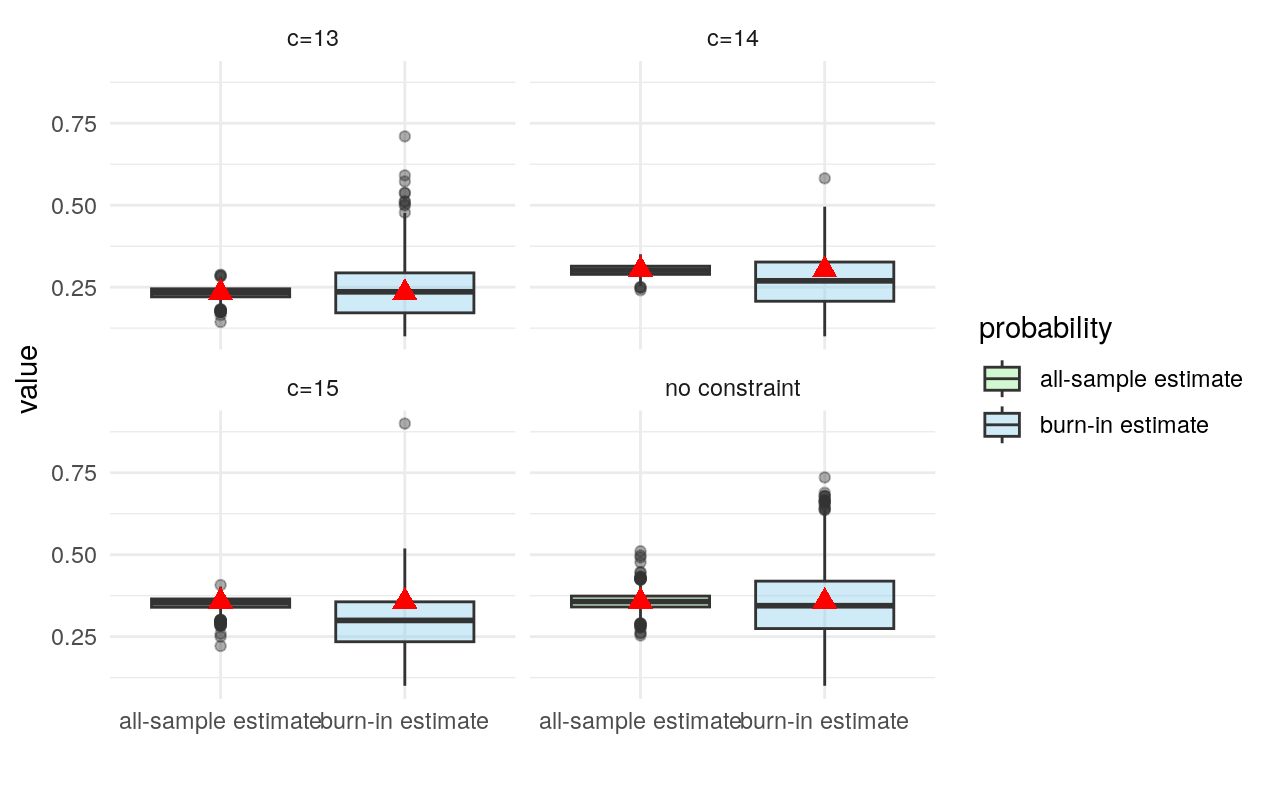}
    %width=140mm,height=60mm]{figure/cara.png}
  \end{center}
  \caption{Estimated allocation probability for different constraint level under RAR. The red triangular marks the target values.}\label{Fig::rar}
\end{figure*}

See Figure \ref{Fig::cara} for the results under CARA. The true allocation probabilities decrease as the constraint level decreases, particularly in stratum 2. The optimal allocation probabilities differ across strata, while both the all-sample estimates and burn-in estimates are generally close to the true values. (Similar to Figure \ref{Fig::rar}, the all-sample estimates show lower variances.) This indicates that a burn-in size of 10 works well to provide a start-up estimate in our model and setting.

\begin{figure*}[htb]
  \begin{center}
    \includegraphics[width=150mm,height=90mm]{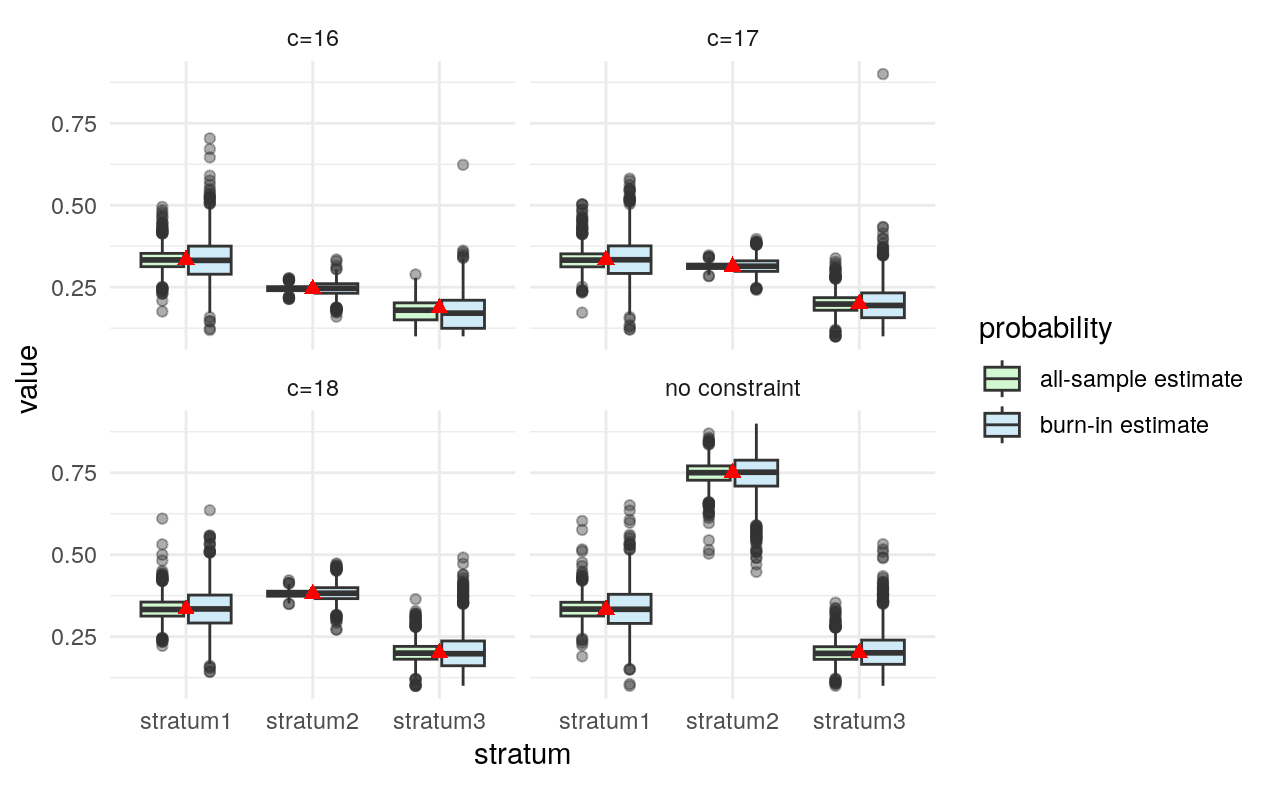}
    %width=140mm,height=60mm]{figure/cara.png}
  \end{center}
  \caption{Estimated allocation probability for different constraint level under CARA. The red triangular marks the true values.}\label{Fig::cara}
\end{figure*}

\subsection{Boxplots for Tables \ref{tab1} and \ref{tab2} \label{secD4}}

In this subsection, we generate boxplots corresponding to the results in Tables \ref{tab1} and \ref{tab2} to provide additional visual comparisons. Specifically, Figure \ref{fig::tab1} corresponds to the data presented in Table \ref{tab1}, while Figure \ref{fig::tab2} correspond to the data in Table \ref{tab2}. However, we found that boxplots are not as effective for comparing variances in our setting, as they do not offer the level of clarity required to illustrate the differences. Therefore, we retain Tables \ref{tab1} and \ref{tab2} in the main text, as they provide a more precise and comprehensive comparison of the variances.

\begin{figure*}[htb]
  \begin{center}
    \includegraphics[width=100mm,height=60mm]{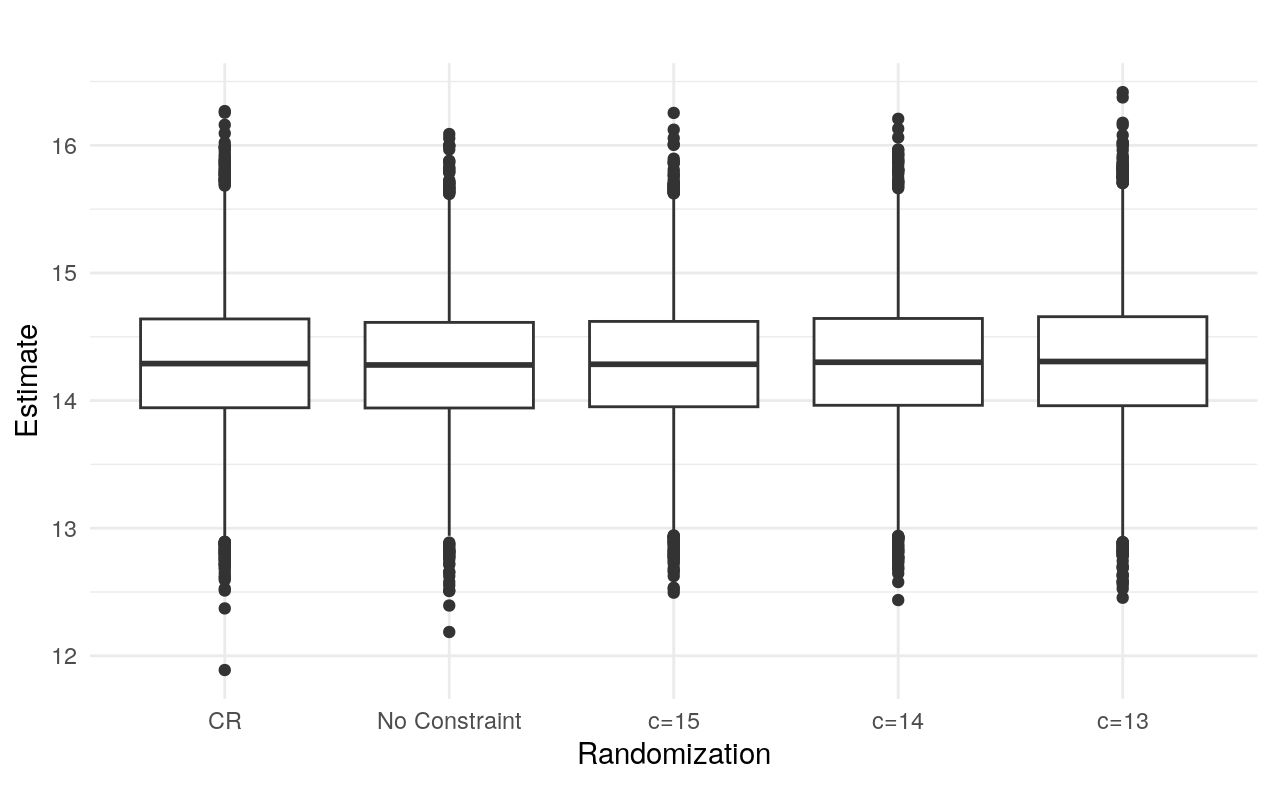}
  \end{center}
  \caption{Boxplots of difference-in-means under different randomization methods (RAR vs. CR).}
  \label{fig::tab1}
\end{figure*}

\begin{figure*}[htb]
  \begin{center}
    \begin{subfigure}[b]{0.7\textwidth}
      \includegraphics[width=\textwidth]{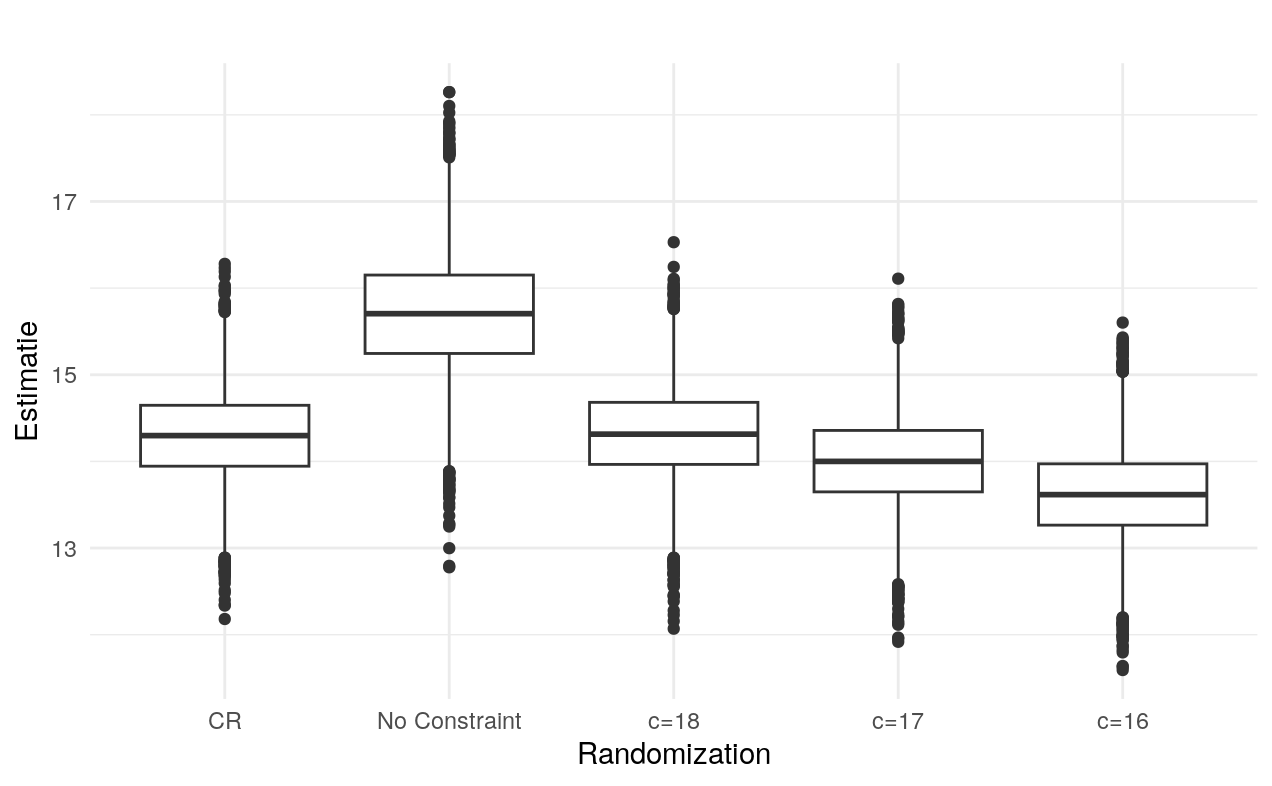}
      \caption{Boxplots of difference-in-means.}
    \end{subfigure}
    
    \vspace{10pt} % Vertical space between subfigures
    
    \begin{subfigure}[b]{0.7\textwidth}
      \includegraphics[width=\textwidth]{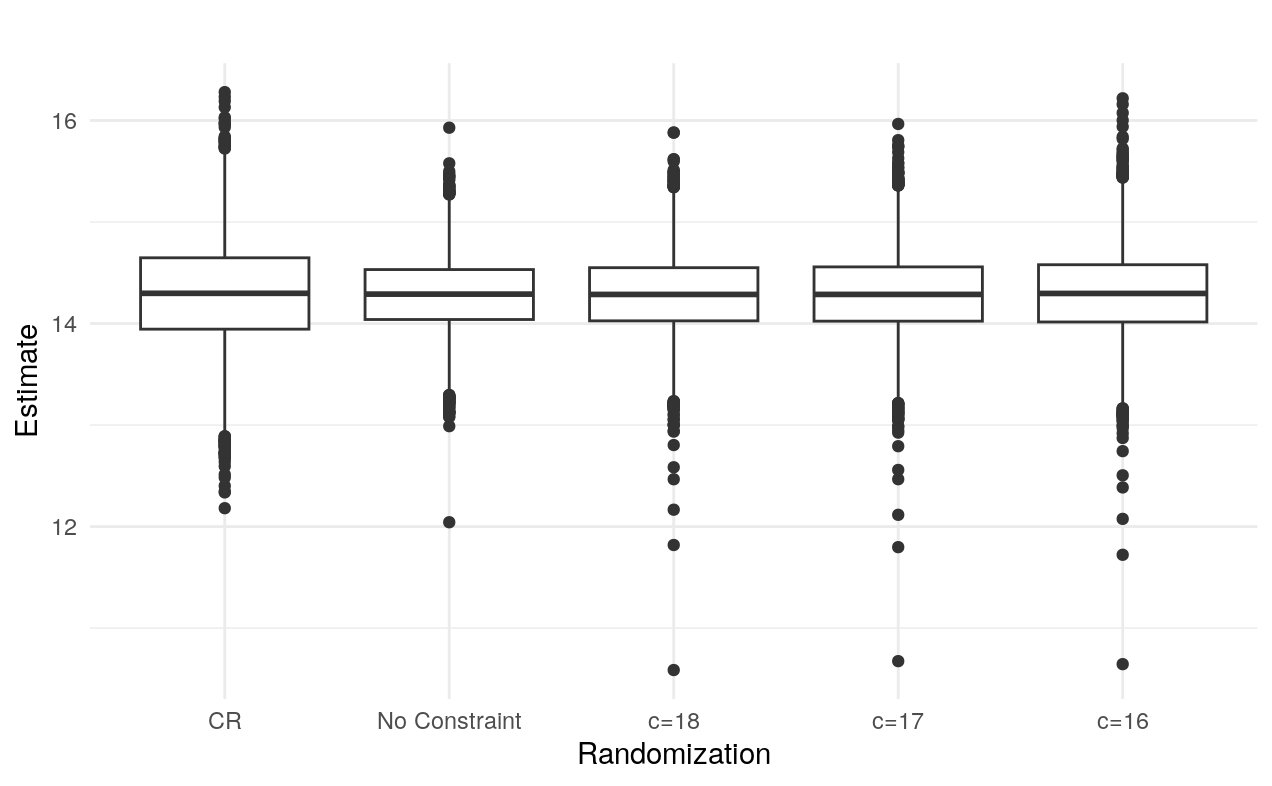}
      \caption{Boxplots of stratified difference-in-means.}
      
    \end{subfigure}
    
    \caption{Boxplots of difference-in-means and stratified difference-in-means under different randomization methods (CARA vs. CR).}
    \label{fig::tab2}
  \end{center}
\end{figure*}

\end{document}